\documentclass[paper]{JHEP3} \pdfoutput=1 
\usepackage{amsmath}
\usepackage{url}
\usepackage[english]{babel}

\usepackage{cite}
\usepackage{amsmath}

\usepackage{epsfig}

\newcommand{\tmop}[1]{\ensuremath{\operatorname{#1}}}

\newcommand\sss{\mathchoice%
{\displaystyle}%
{\scriptstyle}%
{\scriptscriptstyle}%
{\scriptscriptstyle}%
}

\newcommand\LambdaQCD{\Lambda_{\rm\scriptscriptstyle QCD}}

\newcommand\mathd{\mathrm{d}}

\newcommand\PhiB{\Phi_{\rm \scriptscriptstyle B}}
\newcommand\PhiR{\Phi_{\rm r}}

\newcommand\as{\alpha_{\sss\rm S}}

\newcommand\kt{k_{\sss\rm T}}
\newcommand\pt{p_{\sss\rm T}}

\newcommand\fastjet{{\tt FastJet}}
\newcommand\MCatNLO{{\tt MC@NLO}}
\newcommand\mcfm{{\tt MCFM}}
\newcommand\POWHEGBOX{{\tt POWHEG BOX}}
\newcommand\POWHEG{{\tt POWHEG}}
\newcommand\scalup{{\tt scalup}}
\newcommand\POWHEGBOXVtwo{{\tt POWHEG BOX V2}}

\newcommand\Vtwo{{\tt V2}}
\newcommand\PYTHIA{{\tt PYTHIA}}
\newcommand\PYTHIAEight{{\tt PYTHIA8}}

\newcommand\HERWIG{{\tt HERWIG}}

\newcommand\MadGraphFive{{\tt MadGraph5}}

\def\({\left(} 
\def\){\right)} 

\def\beq{\begin{equation}}
\def\beqn{\begin{eqnarray}}
\def\eeq{\end{equation}}
\def\eeqn{\end{eqnarray}}

\title{Top-pair production and decay at NLO matched with parton showers}
\vfill
\author{John M. Campbell\\ Fermilab, Batavia IL 60510, USA\\
E-mail: \email{johnmc@fnal.gov}}
\author{R. Keith Ellis\\ Fermilab, Batavia IL 60510, USA\\
E-mail: \email{ellis@fnal.gov}}

\author{Paolo Nason \\
INFN, Sezione di Milano Bicocca, Italy\\
E-mail: \email{Paolo.Nason@mib.infn.it}}
\author{Emanuele Re \\
Rudolf Peierls Centre for Theoretical Physics, 1 Keble Road, University of Oxford, UK\\
E-mail: \email{emanuele.re@physics.ox.ac.uk}}

\keywords{POWHEG, SMC, NLO, QCD}

\abstract{
We present a next-to-leading order (NLO) calculation of $t\bar{t}$
production in hadronic collisions interfaced to shower generators
according to the \POWHEG{} method. %
We start from an NLO result from previous work, obtained in the zero width
limit, where radiative corrections to both production and decays are included.
The \POWHEG{} interface required an extension of the \POWHEGBOX{} framework,
in order to deal with radiation from the decay of resonances. This extension
is fully general (\emph{i.e.} it can be applied in principle to any process
considered in the zero width limit), and is here applied for the first time.
In order to perform a realistic simulation, we
introduce finite width effects using different approximations, that we
validated by comparing with published exact NLO results.
We have interfaced our \POWHEG{} code to the \PYTHIAEight{} shower Monte Carlo
generator. At this stage, we dealt with novel issues related to the
treatment of resonances, especially
with regard to the initial scale for the shower that needs to be set
appropriately.
This procedure affects, for example, the fragmentation function of the $b$ quark,
that we have studied with particular attention.
We believe that the tool presented here improves over previous generators for all
aspects that have to do with top decays, and especially for the study of issues
related to top mass measurements that involve $B$ hadrons or $b$ jets.
The work presented here also constitutes a first step towards a fully consistent
matching of NLO calculations involving intermediate
resonances decaying into coloured particles, with parton showers.
}

\preprint{FERMILAB-PUB-14-504-T\\
OUTP-14-17P}

\begin{document}
\section{Introduction}
Top production at the LHC is reaching unprecedented levels of
precision. The large number of analyzed $t\bar{t}$ events has allowed
ATLAS and CMS to perform accurate measurements of the production cross
section
\cite{Chatrchyan:2013ual,Chatrchyan:2012ria,Chatrchyan:2013faa,
  Aad:2014iaa,Aad:2014kva,Aad:2012vip},
of the top mass
\cite{Chatrchyan:2013xza,Chatrchyan:2013haa,Chatrchyan:2013boa},
and of the differential distributions of the produced pairs
\cite{Chatrchyan:2013wua,Chatrchyan:2012saa,Aad:2014zka,Aad:2012hg}.

Top quark pair production is now known at NNLO for total
rates~\cite{Czakon:2013goa}, and complete 
differential distributions can be expected to be computed at NNLO 
in the future too~\cite{Abelof:2014fza,Czakon:2014xsa}, at least for
on-shell top quarks, and possibly also including factorizable NNLO
corrections to the decay~\cite{Brucherseifer:2013iv}.

The differential cross section for the on-shell
production of $t\bar{t}$ pairs at NLO has been available for a long
time~\cite{Mangano:1991jk}, and has been implemented as NLO calculations matched to
a parton shower generator (NLO+PS from now on) in the
\MCatNLO{} framework~\cite{Frixione:2003ei}, and in the \POWHEG{}
framework~\cite{Frixione:2007nw}.
In both these NLO+PS implementations, the top decay is treated
in an approximate way, according to the prescription of ref.~\cite{Frixione:2007zp},
and no radiative corrections to the decay processes are included.

In ref.~\cite{Bernreuther:2004jv,Melnikov:2009dn,Campbell:2012uf}, NLO
radiative corrections to production and decay have been computed. The
top is treated there as an on-shell particle, with production and
decay processes fully factorized. Radiative corrections to $W$
hadronic decays are also included there. Currently, however, the most
accurate description of top-quark pair production with fully-exclusive
decays is the one obtained at NLO in
refs.~\cite{Denner:2010jp,Denner:2012yc}, (that will be referred to in
the rest of this paper as DDKP), and in
refs.~\cite{Bevilacqua:2010qb,Heinrich:2013qaa}. These publications
contain a computation of NLO corrections to the process $pp\to e^+
\nu_e \mu^- \bar{\nu}_\mu b \bar{b}$, hence spin-correlation as well
as intermediate top offshellness effects are taken into account
exactly.\footnote{Leptonic $W$ decays are exact in
  refs.~\cite{Bevilacqua:2010qb,Denner:2012yc} and in the narrow width
  approximation in refs.~\cite{Denner:2010jp,Heinrich:2013qaa}. In
  ref.~\cite{Heinrich:2013qaa} $b$-initiated channels were also
  included.}  More recently, the same computation has been also
performed in the 4-flavour
scheme~\cite{Frederix:2013gra,Cascioli:2013wga}.

\begin{figure}[!htb]
  \begin{center}
    \includegraphics[width=0.48\textwidth]{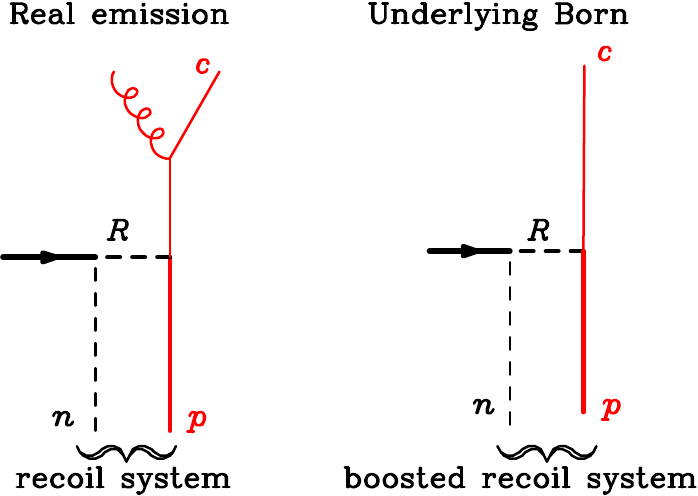}
  \end{center}
  \caption{Example of real emission and underlying Born configuration in a simple model of neutral resonance
    production and decay. The thick solid arrow represents the incoming system. Solid lines represent
    coloured particles, dashed lines represent neutral ones. The particle labelled
    $c$ is the only Born level massless coloured parton in the final
    state. In the standard
    \POWHEGBOX{} framework, the underlying Born is obtained by boosting the $n,p$ system of the real
    emission configuration in order to conserve the total energy in the rest frame of the incoming system.
    With this procedure, the resonance mass is not conserved.}
  \label{fig:modeldec}
\end{figure}

NLO calculations of the complete production process of the top decay products,
also including possibly the hadronic decays of the $W$,
can be directly interfaced to parton
shower generators like \POWHEG{}, as it was actually done in
ref.~\cite{Garzelli:2014dka} using the \POWHEGBOX{} framework. This
procedure, however, is bound to fail in the narrow width limit,
since (in the current \POWHEGBOX{}) both the \POWHEG{} and the subsequent
shower radiation from the resonance decay products
may change the resonance virtuality.
This problem is particularly evident in cases when a strongly decaying
resonance is produced, at the Born level, with non-zero momentum in the partonic CM
frame. This is certainly the case for $t\bar{t}$ production.
In order to discuss this problem in a simpler context,
we consider the production of
a single neutral resonance $R$, decaying into a coloured parton $c$
plus a massive coloured particle $p$, in association
with a neutral particle $n$, as shown in fig.~\ref{fig:modeldec}.
 We also assume colourless incoming particles,
like in an $e^+e^-$ collider, so that we don't need to worry about
initial state radiation. The neutral particle $n$ is included
in such a way that the resonance has non-zero momentum in the CM frame.
The \POWHEG{} formula for this process is
\begin{eqnarray}
\mathd \sigma &=& \mathd\PhiR\, \mathd\PhiB\,
\frac{\bar{B}(\PhiB)}{B(\PhiB)} \times
R(\PhiB,\PhiR)\exp\left[-\int_{\kt^\prime>\kt} \frac{R(\PhiB,\PhiR^\prime)}{B(\PhiB)}
\mathd\PhiR^\prime\right]\,, \nonumber
 \\\label{eq:powhegform}
\bar{B}(\PhiB)&=&B(\PhiB)+V(\PhiB)+\int R(\PhiB,\PhiR) \mathd \PhiR\,,
\end{eqnarray}
where, as usual $B(\PhiB)\mathd \PhiB$ and
 $R(\PhiB,\PhiR) \mathd \PhiR\ \mathd \PhiB$ are respectively the Born and real
cross section, where the real cross section refers to our basic process with the
additional emission of a light parton.
$V$ is the virtual contribution. Its soft and collinear divergences cancel against
those of the real integral in the expression for $\bar{B}$.
Again as usual, the real emission
phase space is factorized in terms of an underlying Born phase space and radiation
variables~\cite{Nason:2004rx}.
The only radiation that we can have now is the one from the decaying
resonance. For this example, in the formalism implemented in prior versions of 
the \POWHEGBOX{}, there is only
one singular region, associated with the radiation of a gluon from
parton $c$. In the corresponding underlying Born, the $c$ parton
momentum is aligned with the momentum of the $c\,g$ system. The direction
of the momentum of the $p\,n$ system in the underlying Born configuration
is the same as in the real emission
configuration, opposite to the $c\,g$ system
(or $c$, in the underlying Born).  The $c$ and $p\,n$ momenta are rescaled
by the same amount, in such a way that the final state energy is the same
in the real configuration and in the underlying Born. The rescaling of the
$p\,n$ momentum is achieved with a boost of the $p\,n$ system along its
direction of flight.
It is now clear that this procedure does not preserve the virtuality of the
resonance. We can easily estimate that the resonance virtuality is changed
by an amount of order $m^2/E$, where $m$ is the virtuality of the $c\,g$
system, and $E$ is its energy. Because of this, we expect distortions
of the radiation distribution when $m^2/E \gtrsim \Gamma$, where $\Gamma$
is the resonance width. It is easy to see what can happen by inspecting
eq.~(\ref{eq:powhegform}). Let us assume that $\PhiB$ is on resonance.
Then $\PhiB,\PhiR$ is on resonance only for $m^2/E \lesssim \Gamma$.
The real integral in $\bar{B}$
is therefore restricted to a vanishing region in the
narrow width limit, thus exposing large negative corrections coming
from the virtual term. This will yield
a large suppression due to the $\bar{B}/B$ ratio.
The Sudakov exponent will yield a substantial suppression only when
$m$ is small enough.
Let us consider now when $\PhiB$ is off resonance.
A potentially large contribution in $\bar{B}$ may be
generated when $\PhiB,\PhiR$ has the resonance on shell, yielding to
a large $\bar{B}/B$ ratio. This should compensate for the reduced
 $\bar{B}/B$ of the previous case. However, in this case the Sudakov exponent
will be quite large, yielding a strong suppression.
We thus expect considerable distortion of the jet mass spectrum
in the region where $m^2/E \simeq \Gamma$.
We anticipate that there shouldn't be problems if $\Gamma$ is large enough,
but it is not easy to quantify this statement. Future studies
comparing the results
obtained with the generator introduced in the present work
with those of ref~\cite{Garzelli:2014dka} will help to clarify this issue.

We remark that problems in the handling of strongly decaying
resonances are not specific to \POWHEG{}. In an \MCatNLO{}
framework~\cite{Frixione:2002ik}, for example, 
the shower itself can change the resonance virtuality.
In order to prevent this one should require that the shower
preserves the resonance mass, and MC counterterms would need to
be modified accordingly.

In this paper, we present an NLO+PS generator for $t\bar{t}$
production that includes NLO corrections in production and decay, in
the narrow width approximation.  In this framework, radiation in
production and in decay can be clearly separated, and no ambiguity
arises for the assignment of the radiated parton, so that we
can avoid the problems mentioned earlier.  As we will show, this
framework will also help us in understanding and dealing with further
problems arising when strongly decaying resonances are involved,
and it may eventually lead to a future consistent
prescription for a framework that fully accounts for finite width and
interference effects.  In order to make our generator fully realistic
and usable, we include an approximate treatment of finite width and
interference effects.

We use the \mcfm{} matrix elements for $t\bar{t}$
production in the narrow width approximation, introduced in
ref.~\cite{Campbell:2012uf} and taken from ref.~\cite{Badger:2011yu}.  
The matrix elements are interfaced to the
\POWHEGBOXVtwo{}, that includes a framework for the treatment of
decaying resonances, including possible radiation from the
decay. The bottom mass is kept finite in our implementation: this
allows us to give a realistic description of bottom fragmentation in
top decays.

The paper is organized as follows. In section~\ref{sec:DescrImp} we
give a description of our generator. In section~\ref{sec:NLOchecks} we
compare our generator at the NLO level with the calculation of
DDKP. The purpose of this comparison is to check
to what extent our calculation maintains its validity outside of the
resonance region, since our implementation of finite width and
interference effects is only approximate.  Finally, in
section~\ref{sec:Pheno} we present our phenomenological studies.
In section~\ref{sec:Conc} we give our conclusion.

\section{Description of the implementation}\label{sec:DescrImp}

\subsection{Resonance decay in \POWHEGBOXVtwo{}}
The \POWHEGBOXVtwo{} framework includes the possibility of
implementing radiative corrections to both production and resonance
decays.  Nested resonances are correctly handled with no limitations.
The implementation is fully automatic as far as the NLO calculation
and its interface to a shower is concerned, consistent with the
general aim of the \POWHEGBOX{} package. It requires external input as
far as the Born phase space and the Born, virtual and real matrix
elements. The information needed for the description of the flavour
structure of the matrix elements is extended in \Vtwo{} to include resonance
information. In the case at hand, for example, the flavour list for
the Born subprocess $g g \to (t\to (W^+ \to e^+ \nu_e)\; b)
\;(\bar{t}\to (W^- \to \mu^- \bar{\nu}_\mu)\; \bar{b})$ is
supplemented with an array that specifies the resonance assignments of
the various final state particles:\newline~\newline
\verb!flst_born(1:nlegborn,j)   =[ 0, 0, 6, -6, 24,-24,-11, 12, 13,-14, 5, -5]!\newline
\verb!flst_bornres(1:nlegborn,j)=[ 0, 0, 0,  0,  3,  4,  5,  5,  6,  6, 3,  4]!\newline~\newline
where the \verb!flst_bornres! entry for $t$ and $\bar{t}$ is zero
(they come from the production process), for $W^+$ and $b$ it is 3
(they come from the decay of the top, which is the 3$^{\tmop{rd}}$
entry) for the $e^+$ and the $\nu_e$ it is 5 (they come from the
$W^+$, which is the 5$^{\tmop{th}}$ entry), and so on.  Similar
structures are assigned for the real emission matrix elements. In this
case, also the emitted parton has a resonance assignment, specifying
whether it originates from the production process, from a top, or from
a $W$.  In the present version of the \POWHEGBOXVtwo{}, it is always
assumed that the resonance assignment at the Born level is the same
for all subprocesses. On the other hand, the resonance assignment of
the radiated parton in real graphs can vary. When computing a real
contribution arising from a given singular region, the program can
figure out the origin (\emph{i.e.} from which resonance the radiation
arises), and the real emission matrix element returns 
the appropriate value on that basis.
All other features, like the generation of an appropriate real radiation
phase space, the soft and collinear limits and the integral of the
soft corrections in case of resonance decays, are handled internally
by the \POWHEGBOXVtwo{}. A full description of these features will
appear in a forthcoming publication, whose draft
is publicly available in the
\verb!Docs/V2-paper.pdf! file in the \POWHEGBOXVtwo{} directory.

The kinematics for the mapping of the real
phase space to an underlying Born configuration departs from the one
described in the \POWHEGBOX{}~\cite{Alioli:2010xd} when
radiation from resonance decays are concerned. In this case,
the mapping is similar to the standard
\POWHEGBOX{} final state mapping, except that the
resonance four-momentum is preserved, rather than the momentum of the
whole final state. For consistency, the integrated soft counterterms
involving the resonance decay are also computed in the resonance frame.  If
the resonance is coloured, it is also included as the initial state particle in
the computation of the corresponding eikonal factors. The soft corrections,
arising from the production process, are instead computed treating the
resonances as final state particles.

\subsection{NLO matrix elements}
As already mentioned in the introduction, the NLO matrix elements are
taken from \mcfm{}~\cite{Campbell:2012uf}. In this calculation, the real contributions
due to radiation in production and radiation in decay are separated, and the
\POWHEG{} subroutine providing the real matrix element selects the appropriate
process, depending upon the origin of the emitting parton.

As illustrated in ref.~\cite{Campbell:2012uf}, in the MCFM program no
radiative corrections to the total width of the top in the decay $t\to W b$ appear.
In fact, such a correction enters the virtual contribution in the top decay, but it
should also enter the total top width appearing in the top propagator.
If the total top width is set to its tree level
value (\emph{i.e.} no strong corrections are included into it), it will exactly cancel the
width for the $t\to W b$ process. In other words, the total cross section
will correspond exactly to the cross section for the production of a $t\bar{t}$
pair, since the branching ratio for $t\to W b$ is one.

The MCFM calculation also computes the width for the W decay. In case of leptonic
decays, this width does not receive strong corrections. The MCFM program thus computes
its tree level value. If the total $W$ width that appears in the $W$ propagators
is an accurate expression including up to fourth order
radiative corrections to hadronic $W$ decays,
the corresponding leptonic branching fraction will have the same accuracy.

As far as hadronic $W$ decays are concerned, the MCFM program computes also their
${\cal O}(\as)$ radiative corrections, that correspond to the usual $1+\as/\pi$
factor. Since radiative corrections to the $W$ hadronic width are known with
much better accuracy, we have preferred to modify the standard MCFM behaviour, by
subtracting this $\as/\pi$ term from the virtual corrections.
In this way, we get a total $W$ hadronic width equal to its
tree-level value, $\Gamma_W^{(0)}({\rm hadrons})$.

We use the following formula for the
$W$ hadronic width:
\begin{equation}
\Gamma_W({\rm hadrons}) = \Gamma_W^{(0)}({\rm hadrons})\times\left(1+ a+1.409\, a^2-12.77\, a^3 -80\, a^4\right)\,,
\end{equation}
with $a=\as(M_W)/\pi$,
and set the total $W$ width in the MCFM program to
\begin{equation}
\Gamma_W=\Gamma_W^{(0)}({\rm leptons})+\Gamma_W({\rm hadrons})\;.
\end{equation}
Furthermore, for each hadronic decay of the $W$, we supply the factor
\begin{equation}
\frac{\Gamma_W({\rm hadrons})}{\Gamma_W^{(0)}({\rm hadrons})}\;.
\end{equation}
It is easy to verify that with this prescription we obtain a total cross section that
is equal to the $t\bar{t}$ cross section, and the correct fraction of leptonic
over hadronic events for each $W$, equal to
\begin{equation}
\frac{\Gamma_W^{(0)}({\rm leptons})}{\Gamma_W({\rm hadrons})}\;.
\end{equation}

\subsection{Finite width and interference effects}\label{sec:finitewidth}
We include finite width and interference effects in an approximate way,
as described in the following.

We generate full Born phase space for off-shell production and decay of a $t\bar{t}$
pair, and we set up a procedure for mapping this phase space into an on-shell one,
that will be used for the calculation of the on-shell matrix elements.
We call $p_t$ and $p_{\bar{t}}$ the top and anti-top off-shell four-momenta,
and  $v_t=\sqrt{p_t^2}$, $v_{\bar{t}}=\sqrt{p_{\bar{t}}^2}$ their
virtualities.
The mapping is such that the off-shell top configuration, with top
and anti-top virtualities $v_t,v_{\bar{t}}$, is mapped into an on-shell
configuration with top mass equal to $(v_t+v_{\bar{t}} )/2$.
The mapping of the $W$ four-momentum invariant is performed as follows.
The top decay phase space, together with the $W$ Breit Wigner shape,
has weight proportional to (neglecting the $b$ mass)
\begin{equation}
\frac{v_t^2-v_W^2}{(v_W^2-m_W^2)^2+m_W^2 \Gamma_W^2} d v_W^2\;,
\end{equation}
that integrates to
\begin{equation}
F(v_W,v_t)=\frac{v_t^2-m_W^2}{M_W\Gamma_W} \arctan \frac{v_W^2-M_W^2}{M_W\Gamma_W}
-\frac{1}{2} \log[(v_W^2-M_W^2)^2+M_W^2\Gamma_W^2]\;,
\end{equation}
where the dimensionful normalization of the argument of the logarithm amounts
to an arbitrary integration constant.
We thus introduce the following function, corrected for the finite $b$ mass
\begin{equation}
\bar{F}(x,y)=F(x,y-m_b)\,,
\end{equation}
and compute the new $W$ virtuality $v_W'$ in such a way that:
\begin{equation}
\frac{{\bar F}(v_W',m_t')-{\bar F}(0,m_t')}{{\bar F}(m_t',m_t')-{\bar F}(0,m_t')}=
\frac{{\bar F}(v_W,v_t)-{\bar F}(0,v_t)}{{\bar F}(v_t,v_t)-{\bar F}(0,v_t)}\;,
\end{equation}
where $m_t'=(v_t+v_{\bar{t}})/2$ is the value of the top mass that will be adopted
in the computation of the matrix elements.
With this choice, the range of the $W$ virtuality remains correctly mapped
between zero and the top mass adopted, and the $W$ is pushed off resonance
only when absolutely necessary.

We now must map the phase space with $v_t$, $v_{\bar t}$, $v_{W^+}$ and $v_{W^-}$
virtualities of the decaying resonances into one with virtualities
$m_t'$, $m_{\bar t}'= m_t'$, $v_{W^+}'$ and $v_{W^-}'$.
This mapped phase space is obtained in a straightforward way
by a fully general mapping procedure, that is applied recursively,
starting with the last decaying resonances (\emph{i.e.} the $W$'s) and ending
with the full final state before any resonance decay. In this procedure
the system comprising
the collision final state before any resonance decay, is also
treated as if arising from a fictitious resonance. In the following,
by \emph{direct} decay products of a resonance we mean its decay products
before any further resonance decay. The mapping procedure is as follows:
\begin{itemize}
\item
Starting from the last decaying resonance in the decay chain,
and following recursively the decay chain in the backward direction, for each
resonance $V$, we go to the $V$ rest frame, and rescale all the momenta of
the $V$ decay products by the same factor, in such a way that energy is conserved
with the newly
assigned $V$ virtuality, and with the newly assigned virtualities of the
direct decay products of $V$ (in case some decay products are
resonances themselves).
\item
If any of the direct decay products of $V$ is itself a decaying
resonance $V'$, the rescaling of the $V'$ momentum is accompanied by
an appropriate longitudinal boost of all decay products (direct and
indirect) of $V'$ along the $V'$ direction, performed in the $V$ rest
frame, such that momentum conservation holds in the decay.
\item
The last decaying resonance is
the fictitious resonance comprising the direct products of the hard collision.
Our procedure is applied also in this case, keeping the energy of the
system fixed.
\end{itemize}

The whole \POWHEG{} generation is carried out with the mapped Born phase
space.  Thus, matrix elements are computed with an on-shell phase
space with equal $t$ and $\bar{t}$ masses. Furthermore, the \POWHEG{}
radiation phase space is computed starting from the mapped Born phase
space, and it thus has equal masses for the top and the anti-top. So,
while we have at hand the original Born phase space and the
corresponding mapped one, we end up with a mapped real phase space, and
we need to build an ``unmapped'' phase space out of that.
Furthermore, this must be done taking into account the origin of the
radiated parton (that is to say, if it was emitted from a resonance or
at the production stage), and preserving the virtuality of all
decaying resonances. This is done using the following mapping procedure:
\begin{itemize}
\item
A longitudinal boost ${\mathcal B}_L$ to the longitudinal rest frame
of the $t{\bar t}$ system is applied to the $t{\bar t}$ system and to
all its decay products, followed by transverse boost ${\mathcal B}_T$
to the $t{\bar t}$ rest frame (unless the radiation is from the hard
process, we have ${\mathcal B}_T={\mathbf 1}$).
\item
The same procedure illustrated earlier for the Born phase space
mapping is applied, where now the original resonance masses are
$m_t'$, $m_{\bar t}'= m_t'$, $v_{W^+}'$ and $v_{W^-}'$, and we want to
end up with $v_t$, $v_{\bar t}$, $v_{W^+}$ and $v_{W^-}$.
\item
The boost ${\mathcal B}_L^{-1}{\mathcal B}_T^{-1}$ is applied to the
$t{\bar t}$ system and to all its decay products.
\end{itemize}
We have also implemented an alternative mapping method, that can be activated
in our program by setting the variable \verb!altmap 1! in the
\verb!powheg.input! file. In case of radiation in resonance decays,
instead of simply rescaling the momenta of the emitter and emitted parton,
we consider the emitter-emitted
pair as a whole, and rescale its four momentum. The
individual emitter-emitted particle momenta are obtained by a suitable
longitudinal boost. This alternative scheme maintains a closer relation
between the transverse momentum of radiation in the on-shell and off-shell
phase spaces. In practice, this feature is not set as default, since
we have not found any noticeable effect due to it.

Notice that in this case the radiated parton
may belong to any decaying resonance, and one must take care to assign it to the
appropriate resonance in order to perform the mapping correctly. This
is not a problem in \POWHEG{}, since the mapping between the Born and
full phase space is always dependent upon the emitting parton, that
thus identifies uniquely the resonance from which the radiated parton
is emitted.

The off-shell phase space is used for the NLO level analysis, and for the
generation of the Les Houches event.

Besides phase space mapping, we need to supply further dynamical information
on the distribution of the top and anti-top virtuality. We have considered
two options for this:
\begin{itemize}
\item
Weight all matrix elements with the Breit-Wigner weights of the top and
anti-top virtualities,~\emph{i.e.}
\begin{equation}
  \frac{v_t^2\Gamma_t^2}{
(v_t^2-m_t^2)^2+(v_t^2\Gamma_t/m_t)^2} \times \frac{v_{\bar t}^2\Gamma_t^2 }{
(v_{\bar t}^2-m_t^2)^2+(v_{\bar t}^2\Gamma_t/m_t)^2}\,.
\end{equation}
This behaviour is activated by setting the flag \verb!mockoffshell! equal to 1
in the \verb!powheg.input! file.
\item
Default behaviour: we reweight all matrix elements
by the ratio of the full off-shell cross section for the underlying Born process at hand,
including interference effects, over the on-shell Born matrix elements.
Using this procedure we intend to capture, at least at leading order, all
off-shell and interference effects that we have left out in our calculation.
In order to avoid useless complications, we have computed the exact matrix
elements for all decay processes including identical and different fermions
in fully leptonic decays, semileptonic decays into first or second generation
quarks, and fully hadronic decays into two different quark generations
(\emph{i.e.} we don't include interference effects for decays into
the same generation quarks).
For any required flavour combination, the program maps it to the the closest
one that matches one of the above.
The matrix elements were obtained with the \MadGraphFive{} program~\cite{Alwall:2011uj}.
\end{itemize}

\subsection{Les Houches event generation and interfacing to a shower}\label{sec:LHSH}
Our $t{\bar t}$ generator can be run with different options
controlling the radiation from resonance decays. The simplest case is
to exclude radiation from resonance decay, which is achieved in our
program by setting the variable \verb!nlowhich! to 1 in the
\verb!powheg.input! file.  In this case, our generator is an improvement over
the previous implementations of
refs.~\cite{Frixione:2003ei,Frixione:2007nw} only because it includes
spin correlation in production with the full NLO accuracy. 
The events generated in this way can be passed to any Shower Generator like
\PYTHIA{} and \HERWIG{} that is compliant with the standard Les
Houches interface for user processes~\cite{Boos:2001cv}. Shower Monte Carlo's
treat radiation from resonance decays using their own shower formalism, and,
as far as further radiation from the production process is concerned,
radiation hardness is limited by the value of \verb!scalup!,
that is passed to the
program via the Les Houches interface.
This guarantees that the shower
will not generate any radiation harder than the one generated by
\POWHEG{}. No limitations are instead imposed on the radiation of
decaying resonances.

By setting \verb!nlowhich! to 0, radiation from resonances is also
generated. In this case, \POWHEG{} uses the usual highest bid
mechanism to decide whether a given event has initial state radiation
or radiation from one of the resonances. This implies that the
subsequent shower program should also limit the radiation in resonance
decays in such a way that no radiation whatsoever, both from the
production process and from resonance decays, is generated with a
transverse momentum larger than that of the \POWHEG{} generated
radiation. The Les Houches Interface does not provide for a standard
mechanism to veto radiation in resonance decays. In
appendix~\ref{app:pythiaIF} we will explain in detail the procedure
that we adopted to achieve this goal.

A further problem arises in the present case.
When using the above procedure, for each event, at most one of the decaying
resonances will include an NLO accurate radiation.
Furthermore, since initial state
radiation turns out to be much more frequent
(due to the large center of mass energy
available at the production stage), corrections to decays generated by
\POWHEG{} itself will not take place so often, with the consequence
that the Monte Carlo will be mostly responsible for radiation in
resonance decays.  On the other hand, a better description of top decay
is quite desirable, especially for studies aimed at the determination
of the top mass.

In order to deal with the above mentioned problem, a further option is
included, that is activated when the variable \verb!allrad! is
set to 1. If this is the case, the \POWHEG{} generator departs from
the standard behaviour at the stage of the generation of
radiation. Radiation from each allowed singular configuration is
generated, that is to say initial state radiation, radiation in the
decay of the top and anti-top, and if hadronic decays are required,
also radiation from the corresponding decaying $W$. Normally, only the
hardest radiation is kept. In the modified behaviour, all radiations
are instead implemented, so that one can find on the Les Houches event
structure as many radiated partons as there are resonances decaying
into coloured particles, plus possibly one radiation from the
initial state. The real radiation phase space configurations are all
preserved, and at the end they are merged into a single phase space.
The merging can be easily performed by using only two kinds of operations:
a boost of the $W$ system in the $t$ (or $\bar{t}$) rest frame, in
order to account for radiation in top decays,
and a longitudinal boost composed with a transverse boost applied to
the whole $t\bar{t}$ system, including its decay products, in order to
account for initial state radiation.

\section{NLO results}\label{sec:NLOchecks}
\subsection{Comparison with existing results}
Currently, the most accurate description of top-quark pair production
with fully-exclusive decays is the one obtained at NLO in
refs.~\cite{Bevilacqua:2010qb,Denner:2012yc}. These two publications
contain a computation of NLO corrections to the process $pp\to e^+
\nu_e \mu^- \bar{\nu}_\mu b \bar{b}$, hence spin-correlation effects
as well as intermediate top and $W$ offshellness effects are taken
into account exactly. Since no narrow width approximation is used,
they contain at the amplitude level diagrams which are
single-resonant, or non-resonant. In this respect, they are an
improvement upon the results
of~\cite{Bernreuther:2004jv,Melnikov:2009dn}, which were based on the
narrow-width approximation, and an improvement 
upon the MCFM results that we use here.

The computations in refs.~\cite{Bevilacqua:2010qb,Denner:2012yc} have
been performed in the 5-flavour scheme,~\emph{i.e.} $b$~quarks are
considered massless. More recently, two groups have completed the same
computation in the 4-flavour
scheme~\cite{Frederix:2013gra,Cascioli:2013wga}. This is
phenomenologically important since it allows for the first time a
unified description of the $t\bar{t}$, $Wt$ and ``$b$-quark associated
$\ell^+\ell^-\nu\bar{\nu} $ production'' processes (since the $b$
quarks are massive, the result is finite without explicitly requiring
two $b$-jets in the final state, which is obviously needed in a
5-flavour computation).

Since our procedure includes at LO the exact full matrix elements for
$pp\to e^+ \nu_e \mu^- \bar{\nu}_\mu b \bar{b}$ (with massive $b$
quarks), and we also include an approximate treatment of offshellness
effects at NLO via the rescaling procedure explained in
sec.~\ref{sec:finitewidth}, it is interesting to check how well our
results compare to an exact NLO computation. To this end, we have
performed a comparison (as tuned as possible) with the results
presented in DDKP.

The results presented in this section have been obtained for the 8 TeV
LHC, using the following input parameters:
\begin{eqnarray}
m_t&=&172.0\mbox{ GeV}\,,\hspace{0.5cm}
M_W=80.399\mbox{ GeV}\,,\hspace{0.5cm}
M_Z=91.1876\mbox{ GeV}\,,\nonumber \\
\Gamma_W&=& 2.09974\mbox{ GeV}\,,\hspace{0.5cm}
\Gamma_Z=2.50966\mbox{ GeV}\,.
\end{eqnarray}
Since the calculation of DDKP was performed with zero mass for the $b$
quark, the results shown in this section have been obtained setting
$m_b=0$ in our calculation.

When $m_b=0$ the MadGraph matrix elements used for reweighting lead to
a divergent Born cross section, due to the (non-resonant) $g\to
b\bar{b}$ initial and final state subprocesses. The divergent region
is cut-off in the analysis of DDKP by requiring two $b$ jets above a
certain transverse momentum. Within \POWHEG{}, we need to deal with
this problem also at the level of phase-space generation. To this end
we have used the Born suppression technique introduced in
ref.~\cite{Alioli:2010qp} (see also ref.~\cite{Kardos:2014dua} for a
more detailed discussion).  We use a suppression factor of the form
\begin{equation}
\frac{p_{T,b}^2}{p_{T,b}^2+p_{T,{\rm supp}}^2}\times
\frac{p_{T,\bar{b}}^2}{p_{T,\bar{b}}^2+p_{T,{\rm supp}}^2}\times
\frac{m_{b\bar{b}}^2}{m_{b\bar{b}}^2 +p_{T,{\rm supp}}^2}\;,
\end{equation}
where $m_{b\bar{b}}^2=(p_b+p_{\bar{b}})^2$ and $p_{T,{\rm supp}}=5\;$GeV. We remark that the Born suppression procedure does not
affect the central value of physical distributions: only the corresponding
statistical errors change.

In this section only, we removed all subprocesses with
$b$-quarks in the initial state, since these channels were not
included in the DDKP paper.

The electroweak coupling $\alpha$ is derived from the Fermi constant
in the $G_\mu$-scheme:
\begin{equation}
  G_\mu = 1.16637 \times 10^{-5} \mbox{ GeV}^{-2}\,,\ \ \ \ 
  \alpha=\frac{\sqrt{2}}{\pi} G_\mu M_W^2 \( 1-\frac{M_W^2}{M_Z^2} \)\,.
\end{equation}
In order to match as closely as possible the procedure adopted in
DDKP, we also use a complex-valued expression for
the weak mixing angle when evaluating the off-shell \MadGraphFive{}
matrix elements,~\emph{i.e.} we define
\begin{equation}
  1-\sin^2\theta_W = \(\frac{M_W^2 - i M_W\Gamma_W}{M_Z^2 - i
    M_Z\Gamma_Z}\)^2 \,.
\end{equation}
On-shell (\mcfm{}) matrix elements are instead computed with real-valued
couplings. The use of a complex-valued weak mixing angle leads to
sub-permille differences, which are appreciable at the level of total
cross-sections, and it is needed to bring our LO results in agreement
to those quoted in DDKP at that level of precision.

We use the LO value $\Gamma^{\rm LO}_t=1.4426$ GeV at LO. At NLO we
use $\Gamma^{\rm NLO}_t=1.3167$ GeV in the \MadGraphFive{} matrix
elements used for reweighting, whilst $\Gamma^{\rm LO}_t$ is used to
evaluate all the core (MCFM) NLO amplitudes. The reweighting procedure
described at the end of sec.~\ref{sec:finitewidth} is corrected by an
overall factor $(\Gamma^{\rm NLO}_t / \Gamma^{\rm LO}_t)^2$, which is
needed to keep the computation consistent with the LO result in the
narrow-width limit.
Finite width effects for the $W$ boson are fully included in all
components of our generator (including the computation of the top
width) by default, thus corresponding to the scheme FwW, in the
notation of DDKP.

In DDKP two choices for the renormalization and
factorization scale are discussed for LHC energies:
\begin{eqnarray}
  \mu_{\rm fix} &=& m_t/2 \,, \nonumber \\
  \mu_{\rm dyn} &=& \frac{1}{2}\sqrt{E_{T,t} E_{T,\bar{t}}}\, \label{eq:mudyn},
\end{eqnarray}
where $E_{T,i}$ is the transverse energy defined as $E^2_{T,i}=m_i^2 +
p^2_{T,i}$.  As in DDKP, when using the dynamic
scale defined in eq.~\eqref{eq:mudyn}, we compute the factorization
and renormalization scale strictly as a function of the event
kinematics.\footnote{The default in the \POWHEGBOX{} is instead to compute the
  scales using the underlying Born kinematics. In order to have the
  alternative behaviour required for this comparison, the variables
  {\tt btlscalereal} and {\tt btlscalect} have to be set to 1 in the
  {\tt powheg.input} file. The subroutine {\tt set\_fac\_ren\_scales}
  is coded in such a way that the scales are computed using the
  underlying Born kinematics for Born, virtual, collinear remnants and
  subtraction contributions, while the real phase space is used for
  the real term.}  This also guarantees the subtraction-scheme
independence of the NLO result.

We use the MSTW2008 parton distribution set~\cite{Martin:2009iq} at
LO/NLO order, and for the results presented in this section, we
evaluate the strong coupling constant using the corresponding LHAPDF
subroutine. In the following we will stick to this set. Other pdf
sets, like those of refs.~\cite{Lai:2010vv,Ball:2010de}, can be used
as well, but we are not interested in a pdf comparison in the present
study.

We cluster final state $b$ quarks and gluons with $|\eta|<5$ into jets
using the anti-$k_T$ algorithm~\cite{Cacciari:2008gp}, with $R=0.5$,
as implemented in the \fastjet{} package~\cite{Cacciari:2011ma}.
The cross section and
distributions shown in the following are defined after these cuts:
\begin{eqnarray}
\label{eq:Dennercuts}
p_{T,j_b} &>& 30 \mbox{ GeV}\,, \ \ \ |\eta_{j_b}| < 2.5\,, \ \ \ p_{T,miss} > 20 \mbox{ GeV}\,, \nonumber \\
p_{T,\ell} &>& 20 \mbox{ GeV}\,, \ \ \ |\eta_{\ell}| < 2.5\,,
\end{eqnarray}
where we require two $b$-jets in the final state.

At LO, we obtain perfect agreement with DDKP, as expected:
\begin{eqnarray}
\sigma_{\rm LO,DDKP}(\mu=\mu_{\rm fix}) &=& 1278.20(4) \mbox{ fb}\,, \ \ \ \ \sigma_{\rm LO,PWG}(\mu=\mu_{\rm fix}) = ({1278.44 \pm 0.15 }) \mbox{ fb}\,, \nonumber \\
\sigma_{\rm LO,DDKP}(\mu=\mu_{\rm dyn}) &=& 1141.69(4) \mbox{ fb}\,, \ \ \ \ \sigma_{\rm LO,PWG}(\mu=\mu_{\rm dyn}) = ({1141.76 \pm 0.13 }) \mbox{ fb}\,. \nonumber \\
\end{eqnarray}

\begin{figure}[!htb]
  \begin{center}
    \includegraphics[width=0.48\textwidth]{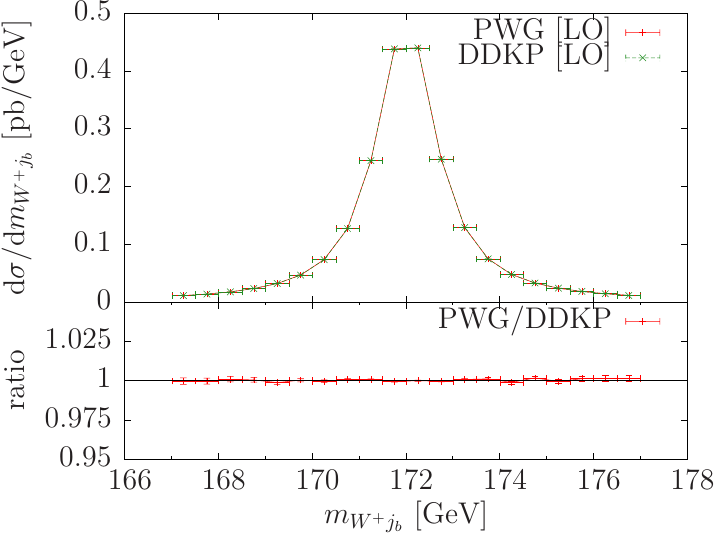}\hspace{0.4cm}
    \includegraphics[width=0.48\textwidth]{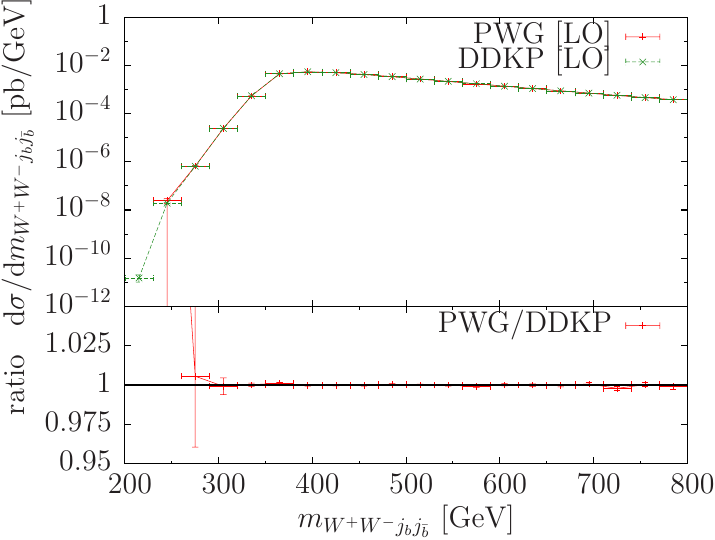}\\
    \includegraphics[width=0.48\textwidth]{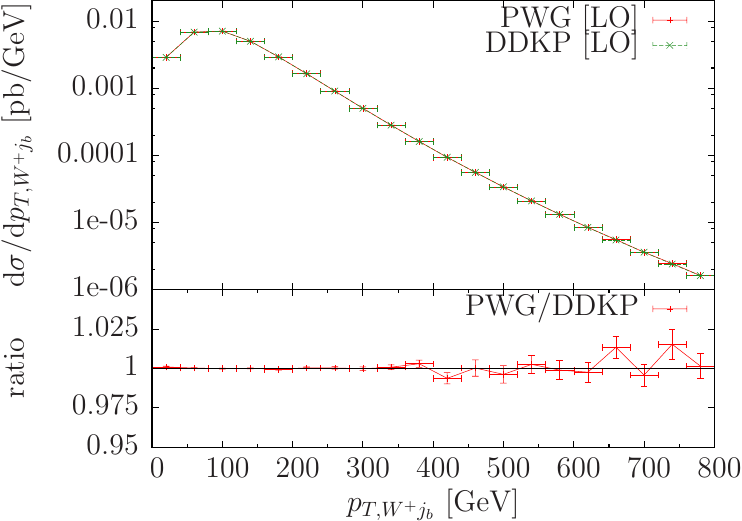}\hspace{0.4cm}
    \includegraphics[width=0.48\textwidth]{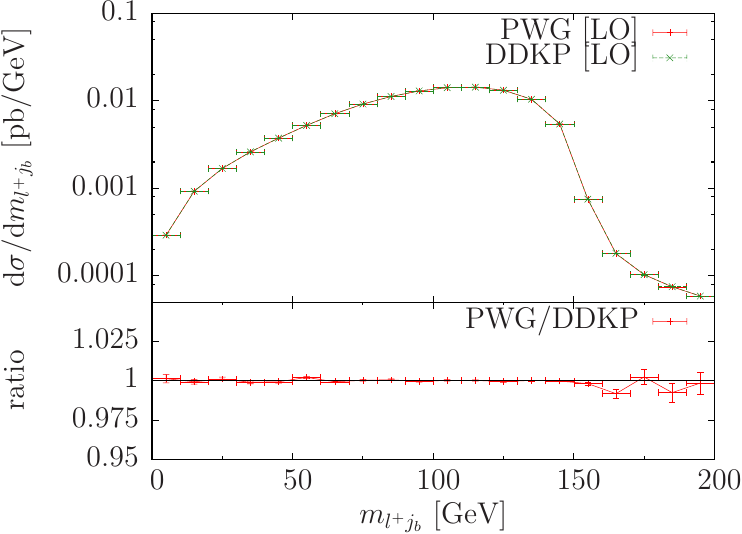}
  \end{center}
  \caption{Comparison between our LO results and those of DDKP,
    obtained using dynamic scales. Plots are obtained with the cuts in
    eq.~\ref{eq:Dennercuts}.}
  \label{fig:LOfull-LODenner}
\end{figure}

Plots in fig.~\ref{fig:LOfull-LODenner} display a comparison between our LO results
and DDKP for a selection of observables, obtained
using the dynamic scale $\mu_{\rm dyn}$.
The observables involving a top or an anti-top shown
in this section are ``reconstructed'',~\emph{i.e.} they are computed
by combining the momenta of the $W$ and $b$-jet associated to the $t$
($\bar{t}$) quark, after the cuts in eq.~\ref{eq:Dennercuts}. All plots
are in good agreement.

At NLO, the procedure we adopted is quite different from the exact
approach used in DDKP. As such, we don't expect
necessarily a perfect agreement for total cross sections, that display
differences of 1$\div$1.5~\%{}:
\begin{eqnarray}
\sigma_{\rm NLO,DDKP}(\mu=\mu_{\rm fix}) &=& 1219(1) \mbox{ fb}\,, \ \ \ \ \sigma_{\rm NLO,PWG}(\mu=\mu_{\rm fix}) = ({1237.3  \pm 0.5})  \mbox{ fb}\,, \nonumber \\
\sigma_{\rm NLO,DDKP}(\mu=\mu_{\rm dyn}) &=& 1221(1) \mbox{ fb}\,, \ \ \ \ \sigma_{\rm NLO,PWG}(\mu=\mu_{\rm dyn}) = ({1209.8  \pm 0.5})  \mbox{ fb}\,. \nonumber \\
\end{eqnarray}

\begin{figure}[!htb]
  \begin{center}
    \includegraphics[width=0.48\textwidth]{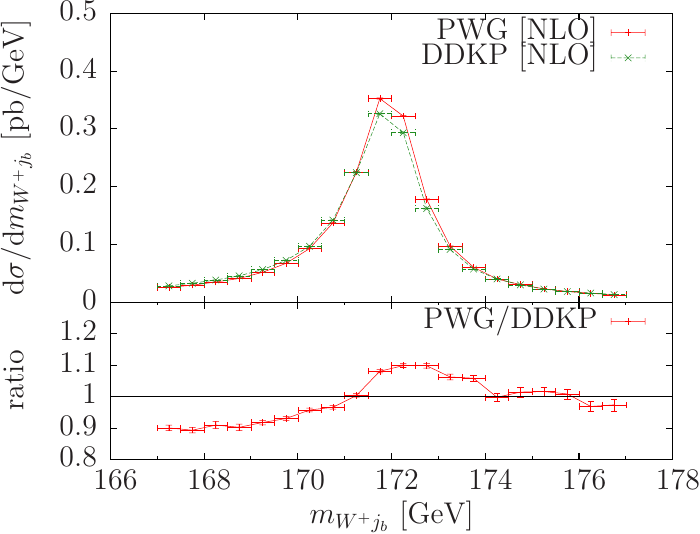}\hspace{0.4cm}
    \includegraphics[width=0.48\textwidth]{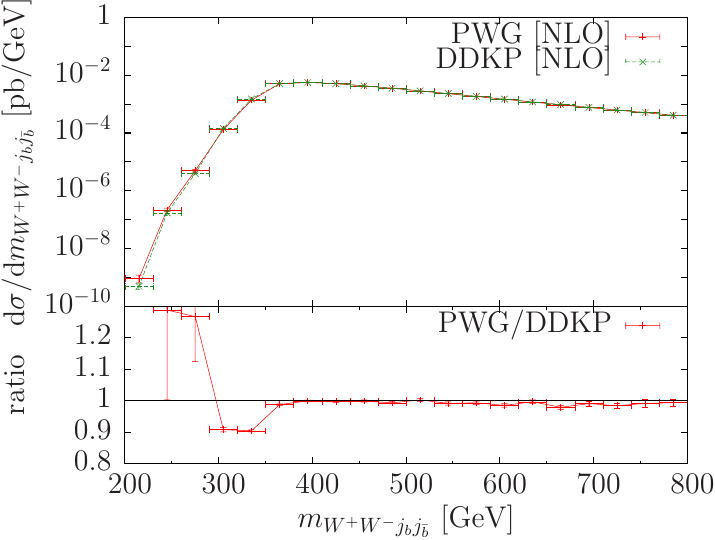}\\
    \includegraphics[width=0.48\textwidth]{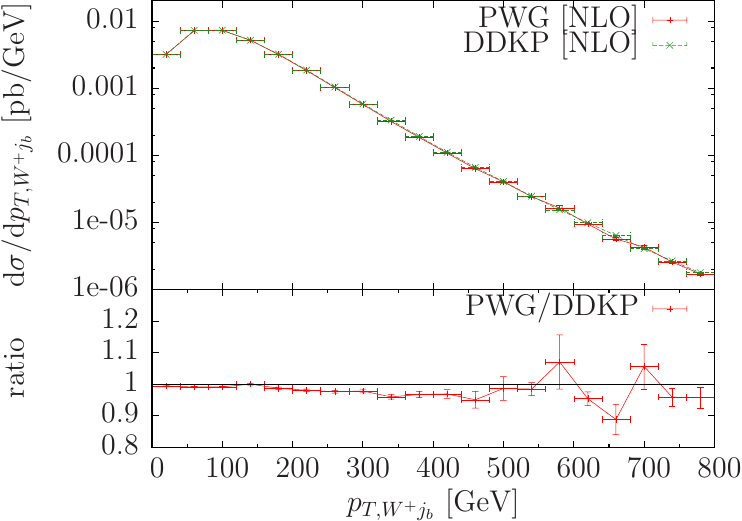}\hspace{0.4cm}
    \includegraphics[width=0.48\textwidth]{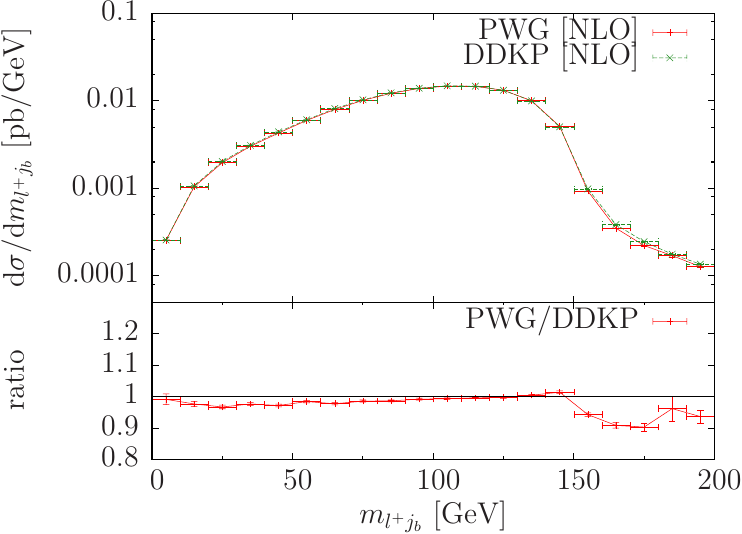}
  \end{center}
  \caption{Comparison between our NLO results and those of DDKP,
    obtained using dynamic scales. Plots are obtained with the cuts in
    eq.~\ref{eq:Dennercuts}.}
  \label{fig:NLOfull-NLODenner}
\end{figure}

In fig.~\ref{fig:NLOfull-NLODenner} we show the same set of
distributions in the NLO case, obtained again with dynamic scales.
A noticeable difference between our NLO result
and the one of DDKP is seen in the shape of the
reconstructed top invariant mass, with our result being about 10\%{} higher
very near the peak. This difference does not appreciably alter the peak
position and width, and is thus unlikely to cause visible errors in,
for example, the determination of the top mass, as can be seen in
the bottom-right plot of the figure.
Small differences are also observed in the $t\bar{t}$ mass in the region
very close to threshold. The lower cross section there is consistent
with our lower cross section for top virtualities below the top mass.
We also see small differences in the transverse momentum distribution of
the top.

We will show in the following that our description
of the peak region turns out to be independent of the
reweighting procedure that we adopt,~\emph{i.e.} reweighting using a
simple Breit Wigner weight, or including only double resonant
graphs in the off-shell matrix elements, would yield the same
discrepancy with the DDKP result. This, together with the fact that
we reproduce exactly the DDKP Born result,  indicates that
the differences that we observe at NLO in the peak region must be
due to the radiative corrections,
that in the DDKP case also include interference effects. Since the
code in DDKP is not public, we cannot investigate this issue further.

\subsection{Modelling of offshellness and interference effects at fixed-order}
We now show a comparison, at the NLO level, of our different
approximations to the finite-width and interference effects. The
purpose of this subsection is to show the level of consistency of the
various approximations we use. We consider:
\begin{itemize}
\item the results obtained by reweighting the cross sections
using a Breit-Wigner function. This is obtained
by activating the flag \verb!mockoffshell! as explained earlier.
The corresponding curves are labelled ``BW'';
\item the results obtained by reweighting using the Double Resonant approximation to the
LO off-shell matrix elements, labelled as ``DR'';
\item the results obtained by reweighting with
 the full LO off-shell matrix elements. The corresponding curves are
 labelled as ``full''.
 \end{itemize}
In this section we use a finite value for the
$b$-quark mass ($m_b=4.75$ GeV), which gives $\Gamma^{\rm
  LO}_t=1.4386$ GeV and $\Gamma^{\rm NLO}_t=1.3145$ GeV as values for
the LO and NLO top width, respectively. Observables
labelled with $t$ and $\bar{t}$ refer to quantities at the
``MC truth'' (i.e. Monte Carlo truth) level.
Notice that in our calculation we can always define an ``MC truth'' level
for the top quarks, that corresponds to the top quarks in the
on-shell calculation, mapped with our procedure into the off-shell
kinematics.

\begin{figure}[!htb]
  \begin{center}
    \includegraphics[width=0.25\textwidth]{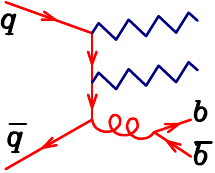}
    \phantom{zzzzzzz}
    \raisebox{0.3cm}{\includegraphics[width=0.25\textwidth]{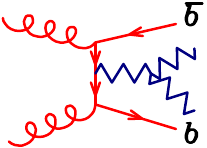}}
  \end{center}
  \caption{Enhanced contributions in the region of small
  invariant mass of the $b\bar{b}$ system (left), and in the
  region of small $b$ transverse momentum (right).}
  \label{fig:enhancedproc}
\end{figure}

Before we begin our comparison, we would like to point out that
there are phase space regions with a relatively large cross section
due to non-resonant contributions: the regions with either of the $b$ quarks
collinear to the beam axis, and the region where the $b$ and $\bar{b}$
quarks are collinear. These regions are kinematically enhanced,
due to the processes represented in fig.~\ref{fig:enhancedproc}.
They can be excluded from the analysis
by imposing a lower limit on the $b$ and $\bar{b}$ transverse momenta (relative
to the beam axis), and on the mass of the $b\bar{b}$ system. Other regions
with potentially important contributions from non-resonant graphs also arise
if we consider same flavour decays of the $W's$, and the corresponding singular
regions should be limited accordingly. Here we limit ourselves to consider
$\mu^+ e^-$ decays, thus avoiding further problems. We remark that
in our calculation all these
non-resonant contributions can only arise from the ``full'' results.

\begin{figure}[!htb]
  \begin{center}
    \includegraphics[width=0.48\textwidth]{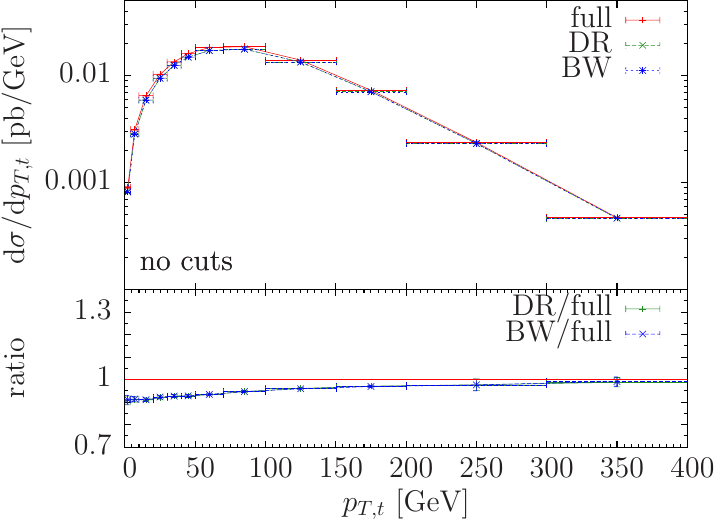}\hspace{0.4cm}
    \includegraphics[width=0.48\textwidth]{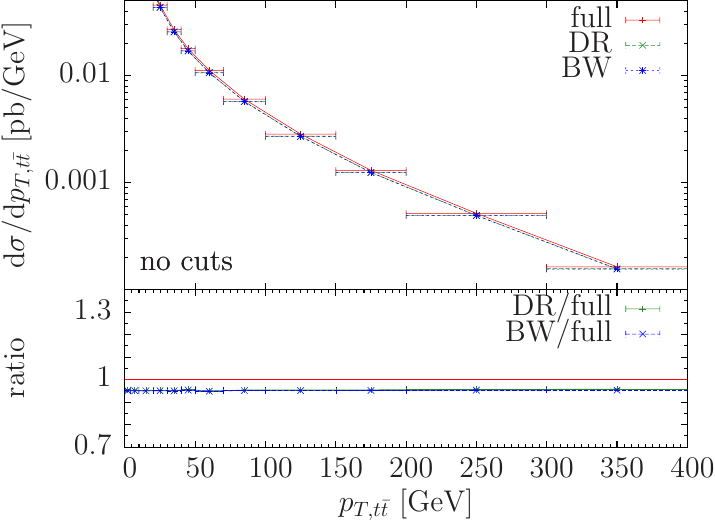}\\
  \end{center}
  \caption{Comparison of our full, DR and BW results for the top and
    for the $t\bar{t}$ pair transverse momenta, without acceptance cuts.}
  \label{fig:inc-NLOfull-NLODR-NLOmock}
\end{figure}

In fig.~\ref{fig:inc-NLOfull-NLODR-NLOmock}
we show our comparisons for inclusive observables that do not depend
strongly upon the off-peak regions. As expected, the BW and DR results
are almost indistinguishable for these observables.  Differences with
respect the full results are visible in the small transverse momentum
region of the top quark, that in fact approaches the threshold region,
thus suppressing the resonant contributions. The full result is
enhanced in this region precisely because of the non-resonant
contributions discussed above.

\begin{figure}[!htb]
  \begin{center}
    \includegraphics[width=0.48\textwidth]{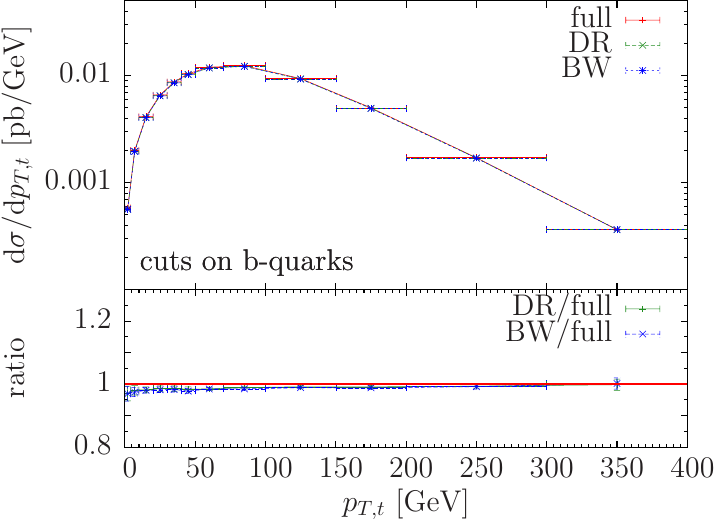}\hspace{0.4cm}
    \includegraphics[width=0.48\textwidth]{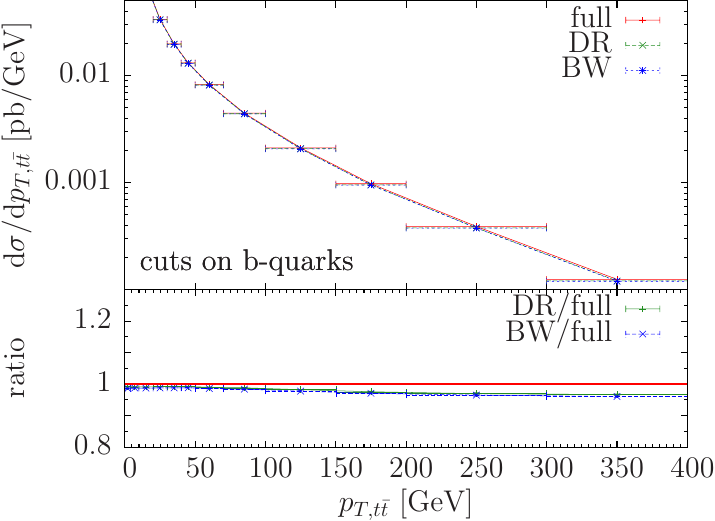}\\
  \end{center}
  \caption{Comparison of our full, DR and BW results for the top and
    for the $t\bar{t}$ pair transverse momenta, imposing the cuts of
    eq.~\ref{eq:sanity-b-cuts}.}
  \label{fig:inc-NLOfull-NLODR-NLOmock-nonrescut}
\end{figure}

In fig.~\ref{fig:inc-NLOfull-NLODR-NLOmock-nonrescut}
we present the same distributions with further cuts to eliminate the
regions with collinear $b$'s,~\emph{i.e.} we require
\begin{equation}\label{eq:sanity-b-cuts}
p_{T,{(b|\bar{b})}}> 30\;{\rm GeV}\,,\quad\quad \sqrt{(p_b+p_{\bar{b}})^2}
> 30\;{\rm GeV}\,.
\end{equation}
These cuts are applied at the parton level, without any jet
definition.  As can be seen, after these cuts are applied, differences
among the various approximations are much smaller. Only few
percent differences are visible in the tail of the transverse momentum
spectrum of the $t\bar{t}$ system.

The effect of the cuts in eq.~\ref{eq:sanity-b-cuts} is better
\begin{figure}[!htb]
  \begin{center}
    \includegraphics[width=0.48\textwidth]{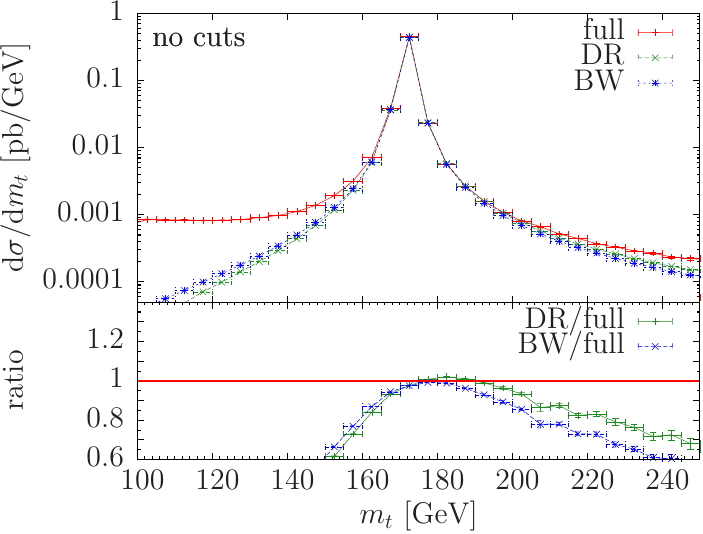}\hspace{0.4cm}
    \includegraphics[width=0.48\textwidth]{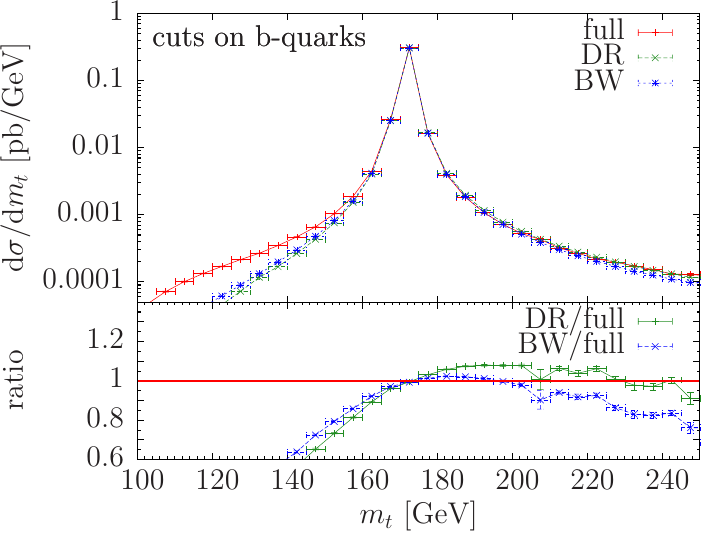}\\
    \includegraphics[width=0.48\textwidth]{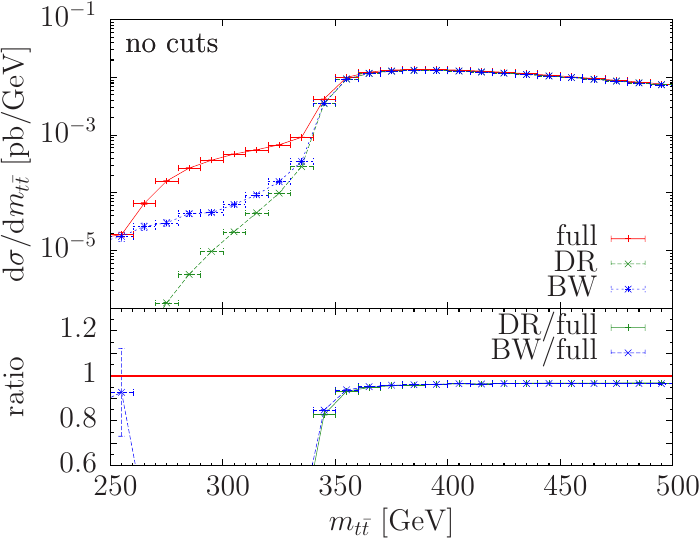}\hspace{0.4cm}
    \includegraphics[width=0.48\textwidth]{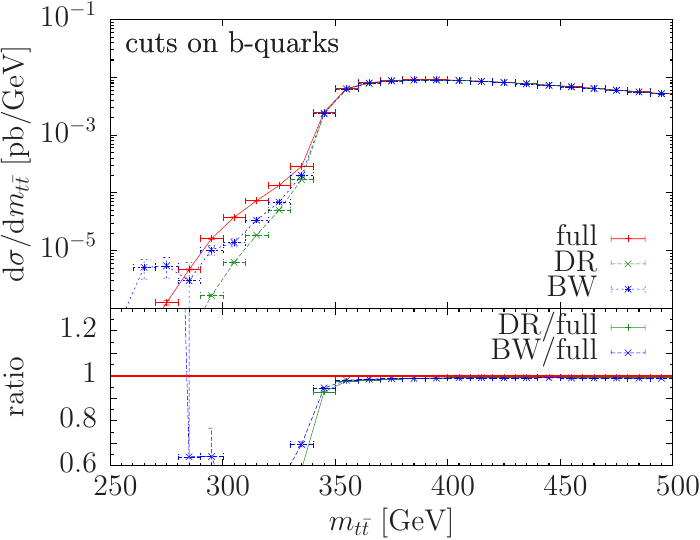}
  \end{center}
  \caption{Comparison of our full, DR and BW results for the top and
    the $t\bar{t}$ system invariant masses, without any cuts (left
    column) and imposing the cuts of
    eq.~\ref{eq:sanity-b-cuts} (right column).}
  \label{fig:mt-mtt-NLOfull-NLODR-NLOmock}
\end{figure}
displayed in fig.~\ref{fig:mt-mtt-NLOfull-NLODR-NLOmock}, where we
compare the invariant mass of the top-quark and $t\bar{t}$ system
(both at the MC truth level). Here we notice that without imposing
the cuts on $b$-quarks, the results from our different approximations
are quite different: below the peak, the $m_t$
distribution in the ``full'' case falls off much less rapidly, due to
the effect of collinearly-enhanced non-resonant contributions, which
affect, to a minor extent, also the tail above the peak. The ``DR''
and ``BW'' results are instead quite similar, especially close to the
peak, as expected. After including the cuts in
eq.~\ref{eq:sanity-b-cuts}, the three predictions are much closer to
each other. Far below the peak, non-resonant contributions are again
noticeable, but the ``full'' distribution now falls off more rapidly,
since collinear-enhanced terms are suppressed by the cuts.  The
results for the invariant mass of the $t\bar{t}$ system show that
there are ambiguities in the final results when approximations like
``DR'' and ``BW'' are used, especially below threshold. After cuts,
all the three curves are closer, except below threshold, where the
``full'' result can be considered our best prediction, although it
misses genuine NLO effects in this region.

Since in this work we are mainly interested in studying phase space
regions close to the peak, in the rest of this section we will
show results after the cuts of eq.~\ref{eq:sanity-b-cuts}.

\begin{figure}[!htb]
  \begin{center}
    \includegraphics[width=0.48\textwidth]{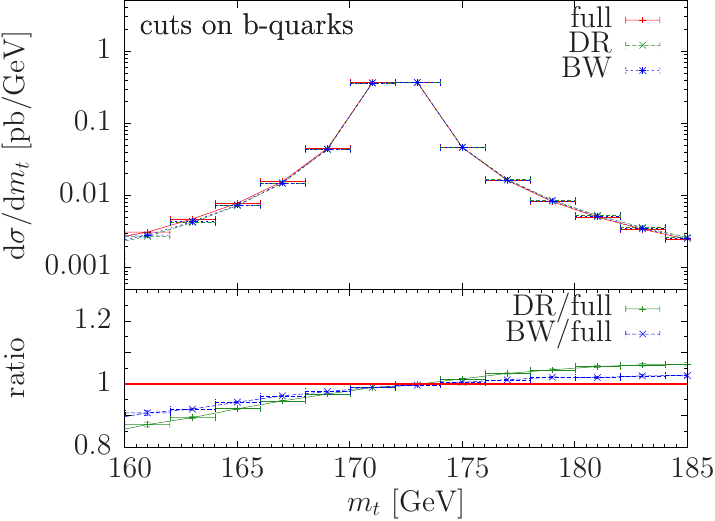}\hspace{0.4cm}
    \includegraphics[width=0.48\textwidth]{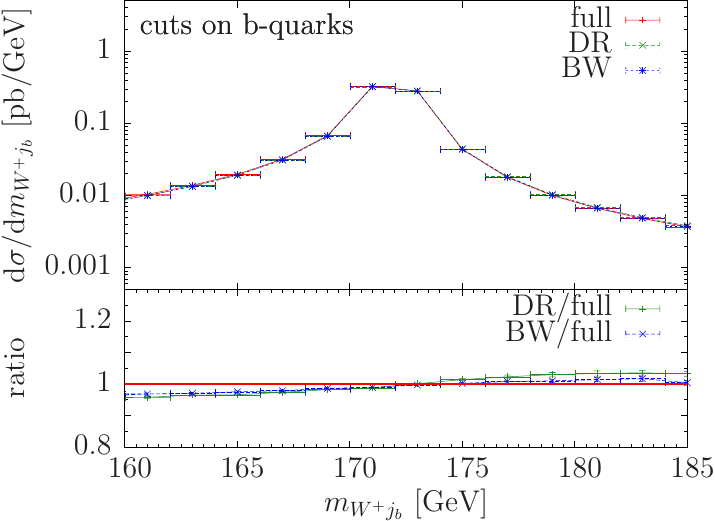}
  \end{center}
  \caption{Comparison of our full, DR and BW results for the top mass
    and the mass of the $W^+$-$b$-jet system, imposing the cuts of
    eq.~\ref{eq:sanity-b-cuts}.}
  \label{fig:peak-NLOfull-NLODR-NLOmock}
\end{figure}

In fig.~\ref{fig:peak-NLOfull-NLODR-NLOmock}
we display our results for the top-quark invariant mass, as defined
from the MC truth, or as reconstructed from the $W^+$ and the $b$
jet. In both cases we see a fair agreement between the BW and DR
results around the peak area, that is worsening as we go above or
below the peak (and more so for the $m_t$ plot).
We notice that the reweighting
applied in DR can compensate unwanted features of
the mapping procedure. This is not the case for the BW result,
where no reweighting is applied. Thus the agreement of the BW result with the
DR one near the peak region means
that our mapping procedure is quite sound.

The full result
displays a larger distance from the DR and the BW result, especially
below the resonance region.  This is understood as being due to other,
non-resonant production mechanisms entering into play, and being
favoured by the growth of the parton luminosities for smaller
invariant mass of the produced system.  Notice that in the $m_{W^+ j_b}$
distribution (right panel of
fig.~\ref{fig:peak-NLOfull-NLODR-NLOmock}), the difference between
full and DR/BW schemes is reduced with respect to the corresponding
result for ``MC truth'' tops
(fig.~\ref{fig:peak-NLOfull-NLODR-NLOmock}, left panel). This is again
expected, because QCD final state radiation outside the $b$ jet cone
leads to a smaller invariant mass of the reconstructed top. This
effect slightly depletes the peak, and raises the region below it,
as can be clearly seen in the figure,
thus hiding the relative importance of non-resonant production
mechanisms.

We notice that the distribution plotted on the right in
fig.~\ref{fig:peak-NLOfull-NLODR-NLOmock} is fairly similar to the
top left plot in fig.~\ref{fig:NLOfull-NLODenner}, where we compare
with the DDKP results. As anticipated earlier, we interpret the
stability of the distribution for our three levels of
approximation as an indication that the differences must have to
do with the full radiative corrections (including also interference
between production and decay) with respect to the
on-shell ones that we apply.

\begin{figure}[!htb]
  \begin{center}
    \includegraphics[width=0.48\textwidth]{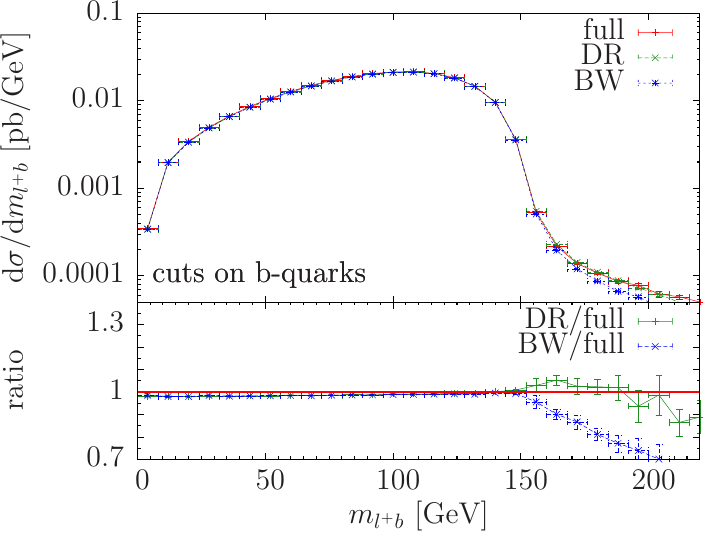}\hspace{0.4cm}
    \includegraphics[width=0.48\textwidth]{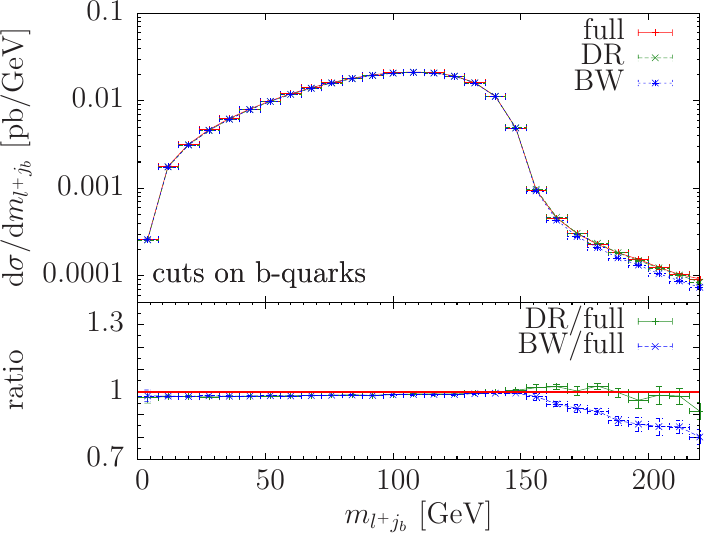}
  \end{center}
  \caption{Comparison of our full, DR and BW results for the mass of
    $l^+$-$b$-quark and $l^+$-$b$-jet systems, imposing the cuts of
    eq.~\ref{eq:sanity-b-cuts}.}
  \label{fig:endpoints-NLOfull-NLODR-NLOmock}
\end{figure}

In fig.~\ref{fig:endpoints-NLOfull-NLODR-NLOmock}
we display the invariant mass distribution of the lepton-$b$-quark and
lepton-$b$-jet systems. As we can see, in realistic conditions, where
we don't have so much control over the neutrino kinematics, the
separation of the three results diminishes, in particular below and near
the end-point region, that is of particular interest for mass
determinations. When larger invariant masses are probed, results start
to deviate, following the same pattern shown in the top right plot of
fig.~\ref{fig:mt-mtt-NLOfull-NLODR-NLOmock}.

\section{\POWHEG{} results}\label{sec:Pheno}
In this section we discuss our \POWHEG{} results, both at the
level of Les Houches (LH) events and after full parton shower (PS).
The observables that we use for our plots are defined as follows:
\begin{itemize}
\item At the LH level $b$, $t$ and $W$ stand for the corresponding LH
  level particles. $b$($\bar{b}$)-jets are obtained clustering all
  coloured final state particles and then selecting those containing
  a $b$($\bar{b}$) quark.
\item at NLO+PS level, $b$ is the $b$-quark after shower
  but before hadronization (we look for the last $b$-quark in the
  shower that has the top as ancestor).  We denote as $B$($\bar{B}$) the
  hardest (\emph{i.e.} largest $\pt$) $b$($\bar{b}$) flavoured hadron,
  irrespective of whether it
  comes from a top or not.  We denote as $j_B$($j_{\bar{B}}$) the jet
  containing the $B$($\bar{B}$).
  The $t$ and $W$ are identified at the MC truth level
  (\emph{i.e.} the last $t$ or $W$ in the shower record).
\end{itemize}

We use the same inputs as those used in the last part of the previous
section, although from here on, unless specified otherwise,
we will impose on all events the same
cuts adopted in DDKP (see eq.~\ref{eq:Dennercuts}),~\emph{i.e.}
\begin{eqnarray}
\label{eq:cuts3}
p_{T,j_B} &>& 30 \mbox{ GeV}\,, \ \ \ |\eta_{j_B}| < 2.5\,, \ \ \ p_{T,miss} > 20 \mbox{ GeV}\,, \nonumber \\
p_{T,\ell} &>& 20 \mbox{ GeV}\,, \ \ \ |\eta_{\ell}| < 2.5\,,
\end{eqnarray}
in the NLO+PS case, and same cuts with $B$ replaced by $b$ in the LH case. 
Jets are reconstructed, as before, with the anti-$\kt$ algorithm
with $R=0.5$.
The renormalization and factorization scales
are set equal to the transverse mass (\emph{i.e.} $\sqrt{m_t^2+p_{T,t}^2}$) of the 
$t$ or $\bar{t}$ at the underlying Born level.

In the following, we will first describe the differences between
results obtained with or without strictly zero widths for the top and the
anti-top. This is to illustrate the need of an off-shell
treatment when analyzing observables useful for the determination
of the top mass. Then we will compare the different approaches that
we have considered for modeling offshellness effects.

\subsection{Comparison of on-shell vs off-shell results}
We first study the differences between our full result, and the
on-shell result, obtained in the strict zero width limit.\footnote{In
this limit interference effects vanish.}
In this section we include radiative corrections both in production and
decay. We do however stick to the usual ``additive'' procedure, that is
to say we do not activate the \verb!allrad! flag discussed in section
\ref{sec:LHSH}. We discuss results at the LH level, and after showering
with \PYTHIAEight{}.

\begin{figure}[!htb]
  \begin{center}
    \includegraphics[width=0.48\textwidth]{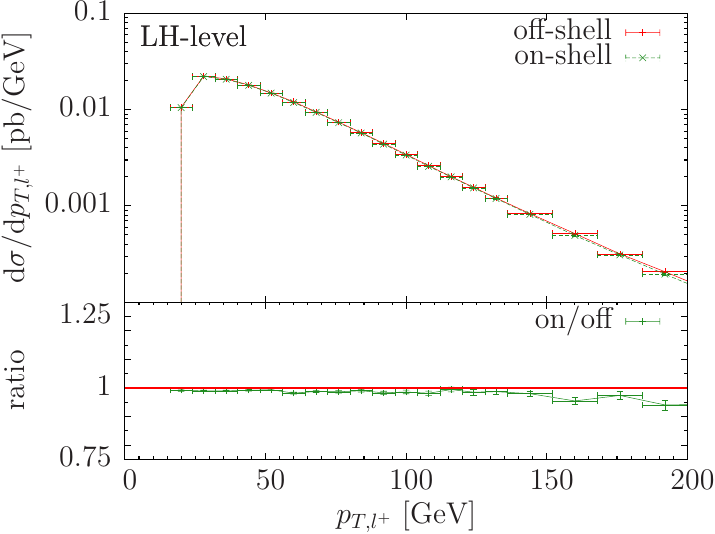}\hspace{0.4cm}
    \includegraphics[width=0.48\textwidth]{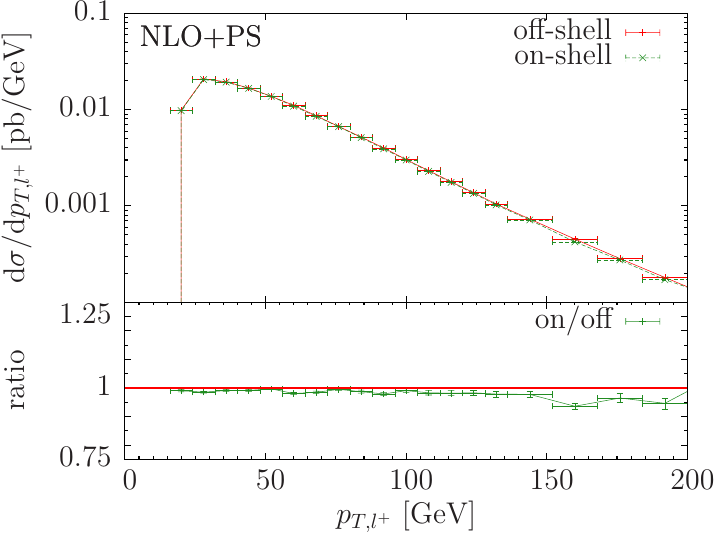}\\
    \includegraphics[width=0.48\textwidth]{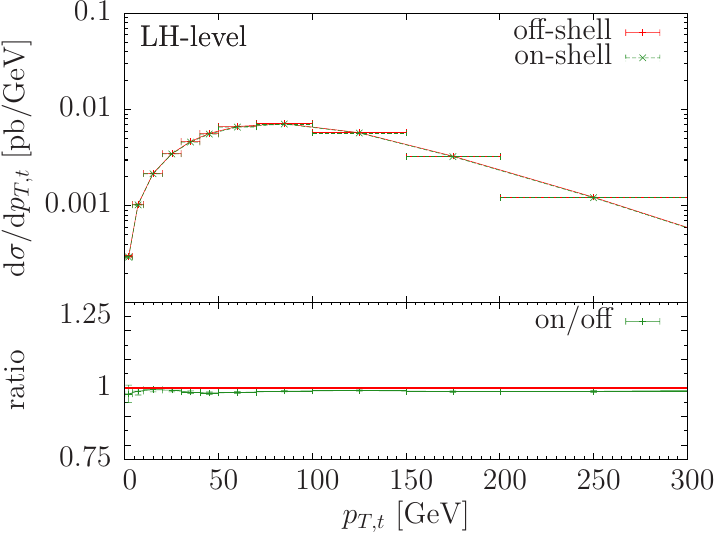}\hspace{0.4cm}
    \includegraphics[width=0.48\textwidth]{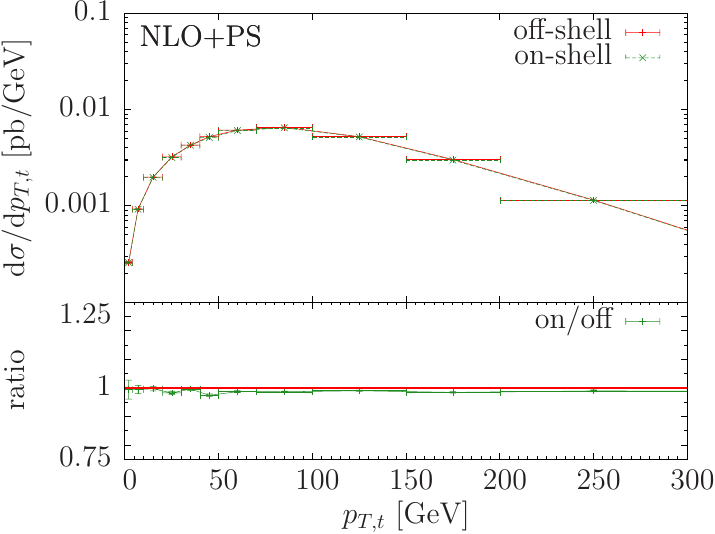}
  \end{center}
  \caption{Comparison of our full result with the results obtained in
    the on-shell approximation, at the hardest-emission level
    (LH-level, left) and after full simulation (NLO+PS, right), using
    the cuts in eq.~\ref{eq:cuts3}.}
  \label{fig:ptl-ptt-off0-on0}
\end{figure}

We begin by showing the lepton and top transverse momentum distributions
in fig.~\ref{fig:ptl-ptt-off0-on0}.
Since the cuts in eq.~\ref{eq:cuts3} remove phase-space regions
where non-resonant diagrams can become collinearly enhanced, we expect
the predictions for these quantities to be in very good
agreement. This is what we observe in the plots, that are shown at the
LH level (left column) and after full parton-showering/hadronization
but without MPI effects with \PYTHIAEight{} (right column).  As
expected, one observes a $K$-factor very close to 1 and constant between the
two predictions.

\begin{figure}[!htb]
  \begin{center}
    \includegraphics[width=0.48\textwidth]{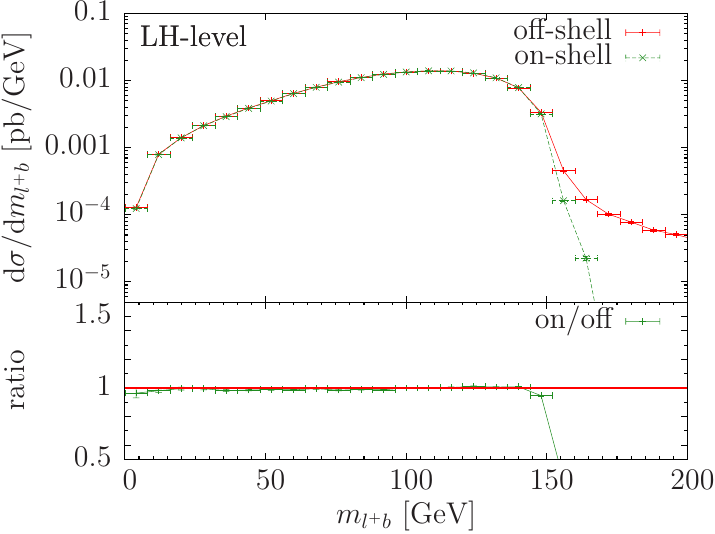}\hspace{0.4cm}
    \includegraphics[width=0.48\textwidth]{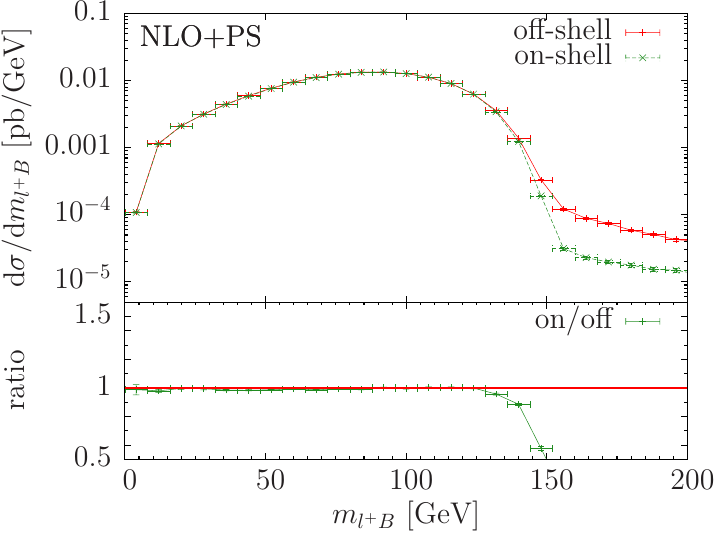}\\
    \includegraphics[width=0.48\textwidth]{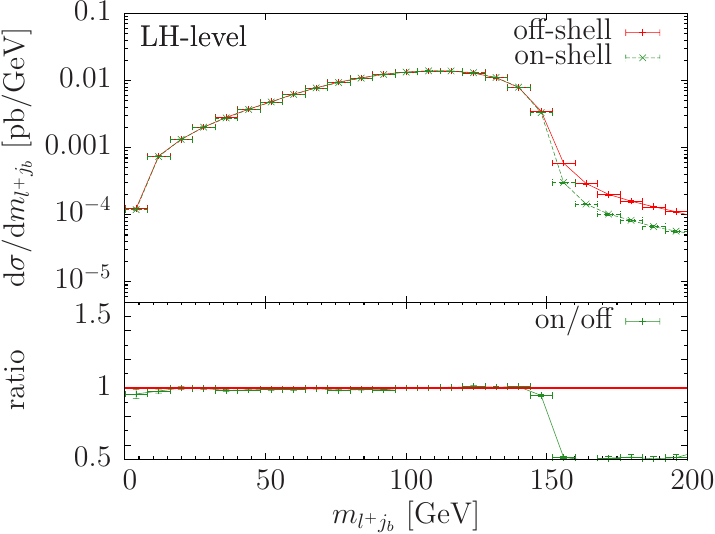}\hspace{0.4cm}
    \includegraphics[width=0.48\textwidth]{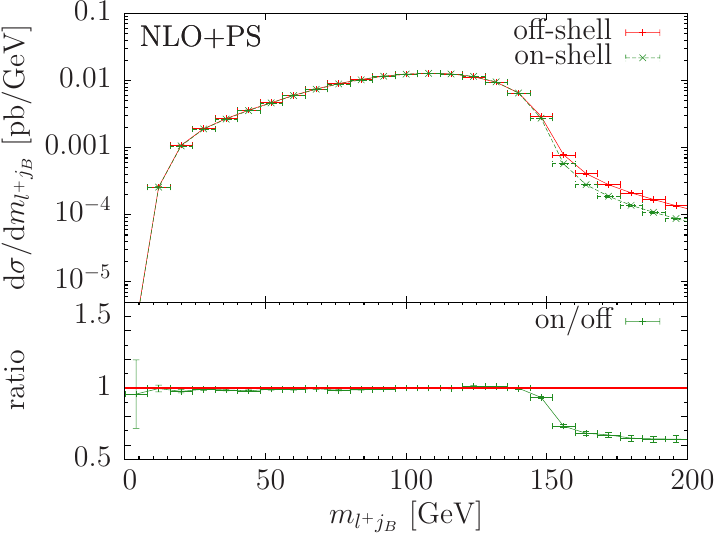}\\
  \end{center}
  \caption{Comparison of our full result with the results obtained in
    the DR approximation, at the hardest-emission level (LH-level,
    left) and after full simulation (NLO+PS, right), using the cuts in
    eq.~\ref{eq:cuts3}.}
  \label{fig:mlb-mlbj-off0-on0}
\end{figure}

In fig.~\ref{fig:mlb-mlbj-off0-on0} we show the invariant mass of the
lepton-$b$ and lepton-$b$-jet systems.
As expected, width effects are particularly relevant near the end-point,
also for the case of the lepton-$b$-jet mass. The effect is larger at the
Les Houches level, where only one radiation can be combined with the
$b$ quark to form a jet. After shower, the larger activity of the event
reduces the effect, that remains however visible.
This is bound to have some impact upon top mass determinations using
end-point observables. Notice that, as expected, in the on-shell calculation
at the Les Houches level there is a sharp threshold in the $b l^+$ invariant
mass. Such sharp threshold is not present in the $l^+ j_b$ case, because initial state radiation
can combine with the $b$ quark to yield a more energetic jet. Also in the $l^+ B$
case we observe a tail above threshold, since colour conservation requires that
the $b$ quark should hadronize in combination with some other
quark not coming from top decay.

\subsection{Impact of different options on off-shell results}
In this section we show a comparison at NLO+PS level between our
results obtained using the traditional ``additive'' \POWHEG{}
procedure (\emph{i.e.} \verb!nlowhich 0!, denoted as ``nlow0''
 in the following plots),  against the
results obtained using the new procedure described at the end of
sec.~\ref{sec:LHSH} (\emph{i.e.}  \verb!nlowhich 0! and \verb!allrad 1!,
denoted as ``allrad'' in the figures).  For
comparison, we also show a curve obtained by setting
\verb!nlowhich 1! 
(denoted as ``nlow1''), that corresponds to let \POWHEG{}
generate only initial state radiation, leaving entirely to the
PS program (\PYTHIAEight{} in our case) the task of emitting gluons from the
$b$-quark originating from the top decay.
We will also compare results with the setting \verb!nlowhich 4! (``nlow4'' in the
figures),
that corresponds to let \POWHEG{}
generate only radiation from top decays (\emph{i.e.}, no  \POWHEG{} radiation
from either production or $\bar{t}$ decays).

\begin{figure}[!htb]
  \begin{center}
    \includegraphics[width=0.48\textwidth]{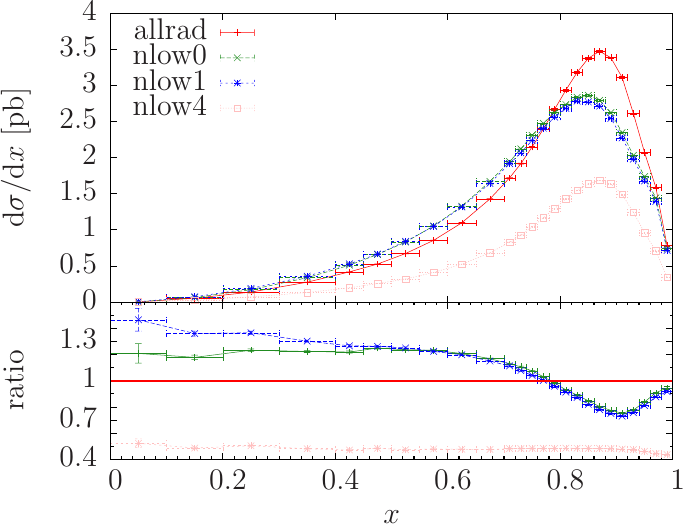}\hspace{0.4cm}
    \includegraphics[width=0.48\textwidth]{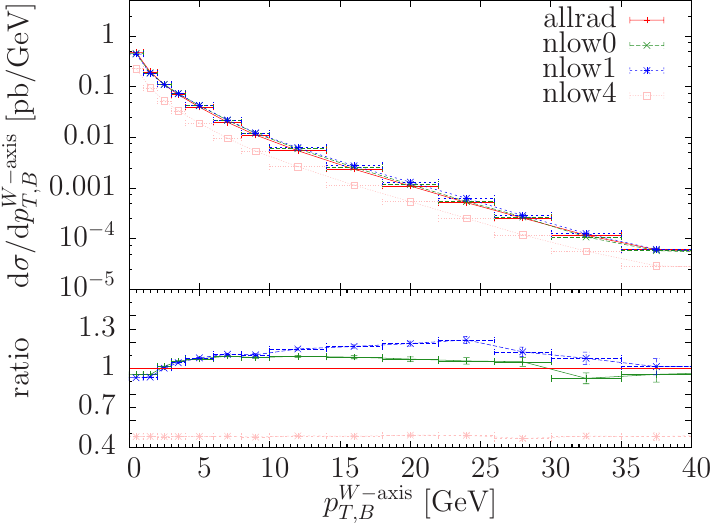}
  \end{center}
  \caption{Properties of the $B$ hadron decay in the top rest frame:
    the $B$ fragmentation function (left), and the $B$ transverse
    momentum with respect to the $W$ direction (right), plotted for
    the four running modes described in the text, and using the cuts
    in eq.~\ref{eq:cuts3}. Notice that in the ``nlow1'' case,
    radiation in decay is fully handled by \PYTHIAEight{} alone.}
  \label{fig:bfrag-allrad-nlow1-nlow0}
\end{figure}

We begin by showing in fig.~\ref{fig:bfrag-allrad-nlow1-nlow0}
two observables that characterize the kinematics of the $B$ hadron coming from
the top decay in the top rest frame,
looking only at the positively charged top:
the $B$ fragmentation function, and the transverse momentum of the
$B$ hadron with respect to the direction of the $W$,
for the four running modes described above.
The $x$ variable of the fragmentation function is defined as the $B$ energy divided by the maximum value that it
can reach at the given $W$ virtuality.  
We immediately see that the ``allrad'' and the ``nlow4'' results
are strictly proportional.
In fact, in both ``allrad'' and ``nlow4'' the hardest radiation in the
top decay is generated by \POWHEG{}, while the subsequent ones are generated
by the parton shower. On the other hand, since the total cross
section in the ``nlow4'' case is equal to the Born cross section,
the two curves differ by the NLO $K$-factor. Thus, the ``allrad'' result compares as
expected with the ``nlow4'', and we now consider the remaining two cases
with respect to the ``allrad'' one.

In the ``nlow1'' case, the radiation in decay is handled by the parton shower alone.
We see that for soft radiation, corresponding to large values of $x$ and small
$p_{T,B}^{W{\rm -axis}}$, the parton shower shape differs appreciably from the
``allrad'' one, and that for hard radiation (corresponding to small values
of $x$ and large $p_{T,B}^{W{\rm -axis}}$) the shower yields a harder result.
We would like to remark, however that the ``allrad'' and the ``nlow4'' results are still quite sensitive
to \PYTHIAEight{} settings. If we turn off the matrix element corrections in \PYTHIAEight{}
(by setting \verb!TimeShower:MEcorrections = off!) our  ``allrad'' result becomes more
similar to the  ``nlow1'' result, being in fact intermediate between the current
 ``allrad'' and  ``nlow1'' curves. In view of the rather elaborate vetoing procedure
that we follow (described in appendix~\ref{app:pythiaIF}), this sensitivity is quite unexpected
and will require more investigation.

In the ``nlow0'' result, \POWHEG{} generates radiation from the production stage
and from $t$ and $\bar{t}$ decay, and it picks the hardest among them.
Initial state radiation is more likely to prevail, because of the wider kinematic
range available, and also because of the stronger colour coupling of the initial
gluons compared to the final $b$ quark. Radiation in top decays prevails only if
relatively hard. This clarifies the behaviour of the ``nlow0'' result in the figure,
that in the hard region approaches the ``allrad'' one, while in the soft
region agrees with the ``nlow1'',~\emph{i.e.} with the parton shower result.
We also notice that the difference between \POWHEG{} and \PYTHIAEight{} in the soft
region, where most of the cross section is, forces a normalization difference
for the distributions in the hard region, since the total normalization
of the curves should be (at least in the absence of cuts) the same.

At this point it is clear that by choosing one of our two most
accurate schemes,~\emph{i.e.} the ``allrad'' or the ``nlow0'', we decide
to rely upon the description of the soft region that is given either by
\POWHEG{}+\PYTHIAEight{}, or (essentially) by \PYTHIAEight{}
alone respectively. Since in the soft region the two frameworks have formally the
same accuracy, more studies are needed to clarify the origin of the
observed difference, that can be due to the used value of
$\LambdaQCD$, to subleading terms in leading log resummation, or to
other reasons. We also notice that because of the interplay of the
soft and non-perturbative regions, the parton shower tune
is also bound to have a non-negligible impact.  We will not perform
such study at this stage. The purpose of the present paper is to
provide a tool by which these questions can be addressed.

\begin{figure}[!htb]
  \begin{center}
    \includegraphics[width=0.48\textwidth]{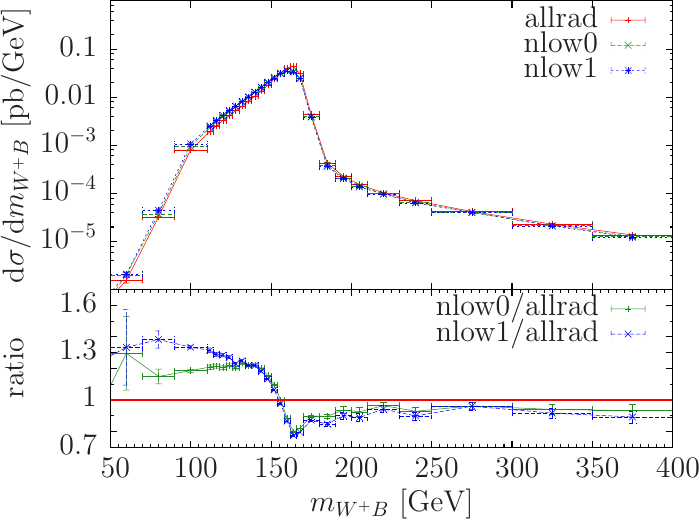}\hspace{0.4cm}
    \includegraphics[width=0.48\textwidth]{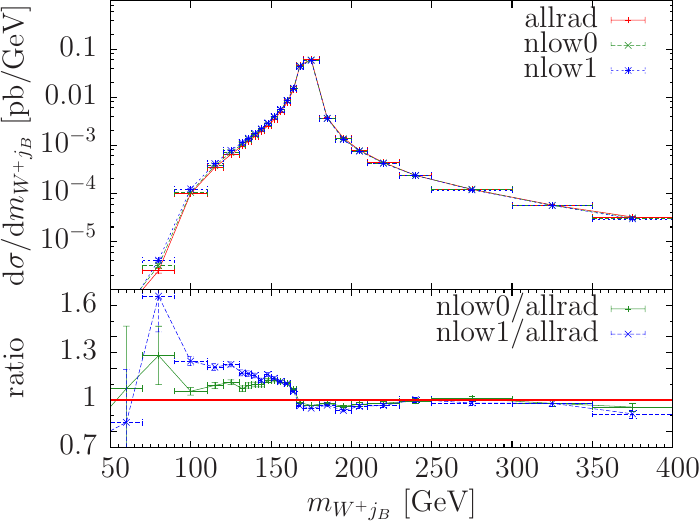}
  \end{center}
  \caption{Mass of the $W$-$B$-hadron and $W$-$B$-jet systems obtained with
    ``nlow1'', ``nlow0'' and ``allrad'', using the cuts in
    eq.~\ref{eq:cuts3}.}
  \label{fig:mwpb-mwpbj-allrad-nlow1-nlow0}
\end{figure}

In figure~\ref{fig:mwpb-mwpbj-allrad-nlow1-nlow0}
we compare the mass of $W^+B$ and $W^+j_B$ systems with ``nlow1'',
``nlow0'' and ``allrad''.  Both these observables should be peaked
near the top mass. However, in the $W^+B$ case, some momentum is lost in
the radiation of the $b$ quark, and we expect that some events from
the top peak will leak to smaller mass values. To a much lesser
extent, this should also happen in the $W^+j_B$ case, since some
radiation is lost out of the jet cone. We see that the differences
between the results displayed in
fig.~\ref{fig:mwpb-mwpbj-allrad-nlow1-nlow0}, can be explained in
terms of this effect, in the light of the results of
fig.~\ref{fig:bfrag-allrad-nlow1-nlow0}.  In fact, the ``nlow1''
result is higher than the ``allrad one'' below the top peak for both
observables, since it has a softer fragmentation function (mostly
affecting the $WB$ observable), and a harder $p_{T,B}^{W{\rm -axis}}$
spectrum (yielding more energy loss outside the jet cone, and thus
affecting the $W^+j_B$ observable). The ``nlow0'' case is similar to
``nlow1'', except that for regions affected by very hard radiation, it
tends to be closer to the ``allrad'' result.

\begin{figure}[!htb]
  \begin{center}
    \includegraphics[width=0.48\textwidth]{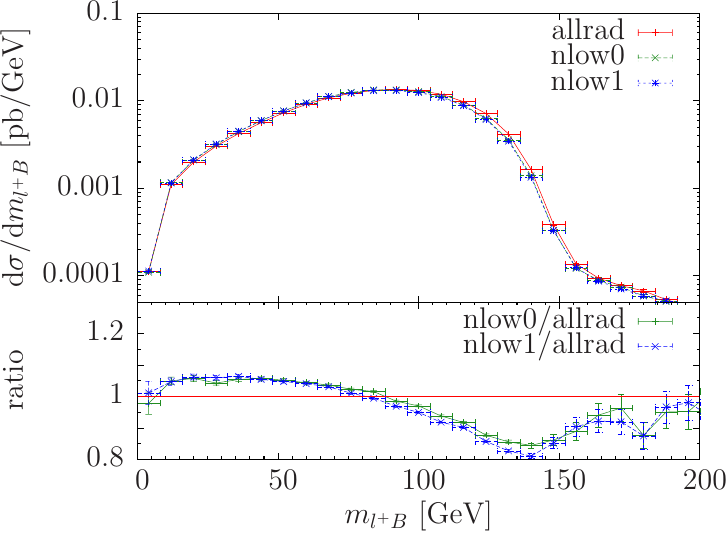}\hspace{0.4cm}
    \includegraphics[width=0.48\textwidth]{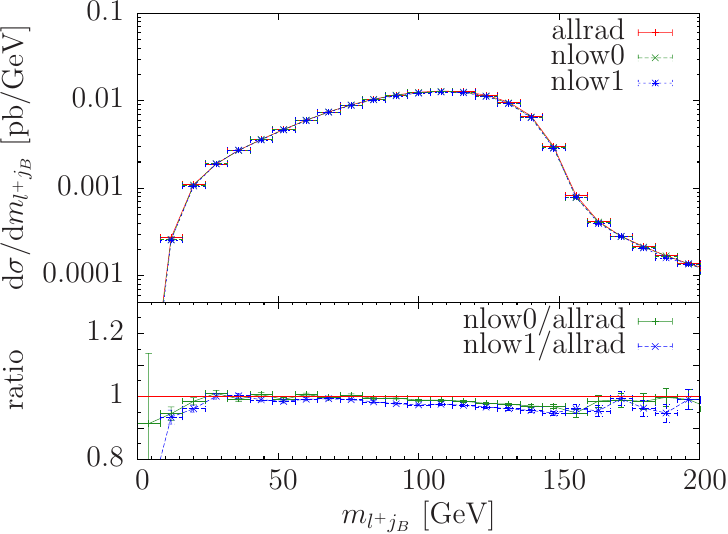}
  \end{center}
  \caption{Mass of the $l^+B$ and $l^+j_B$ systems obtained
    with ``nlow1'', ``nlow0'' and ``allrad'', using the cuts in
    eq.~\ref{eq:cuts3}.}
  \label{fig:mlp-mlbj-allrad-nlow1-nlow0}
\end{figure}

The same effect can be used to explain the differences
observed in the end-point observables $m_{l^+B}$ and $m_{l^+j_B}$,
displayed in fig.~\ref{fig:mlp-mlbj-allrad-nlow1-nlow0}.

\begin{figure}[!htb]
  \begin{center}
    \includegraphics[width=0.48\textwidth]{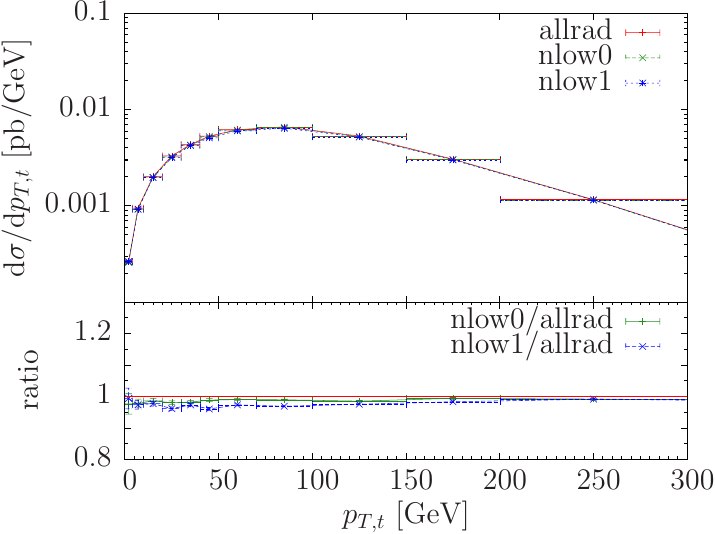}\hspace{0.4cm}
    \includegraphics[width=0.48\textwidth]{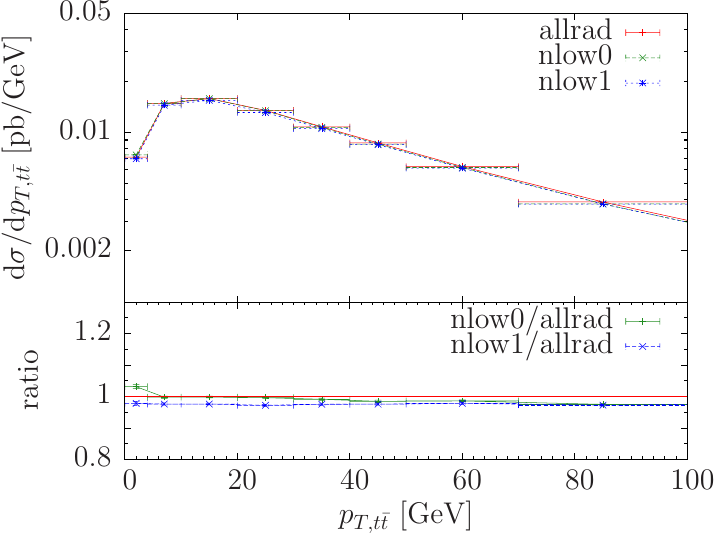}\\
    \includegraphics[width=0.48\textwidth]{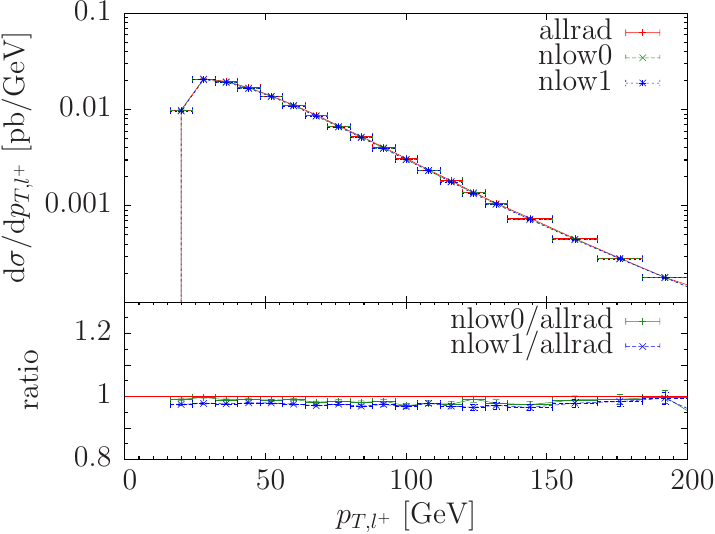}\hspace{0.4cm}
    \includegraphics[width=0.48\textwidth]{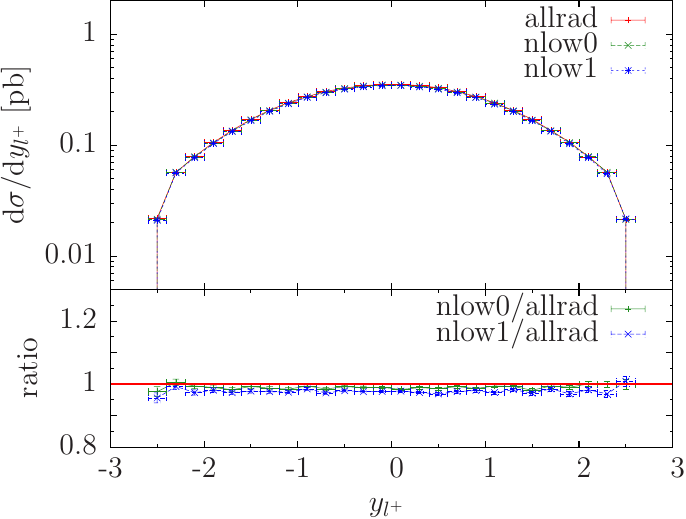}
  \end{center}
  \caption{Transverse momentum of the top and of the $t\bar{t}$ pair
    (top left and right plots), transverse momentum of the lepton
    (bottom left) and rapidity of the lepton (bottom right) obtained
    with ``nlow1'', ``nlow0'' and ``allrad'', using the cuts in
    eq.~\ref{eq:cuts3}.}
  \label{fig:rest-allrad-nlow1-nlow0}
\end{figure}

In fig.~\ref{fig:rest-allrad-nlow1-nlow0}
we display some observables that are not
particularly sensitive to the shape of the top peak.
We see that in this case the three approaches yield quite
similar results. In the transverse momentum of the $t\bar{t}$
pair we observe a difference between the ``nlow0'' and
the other two results. This difference is of little significance,
being about 2.5\%{} in the second bin, and 5\%{} in the first
bin, where however the cross section is rather small.
We have reasons to attribute this difference to an imperfect match
between the meaning of our hardness scales and the ones
adopted by \PYTHIAEight{}. We will discuss further this issue
in appendix~\ref{app:pythiaIF}.

\subsection{Comparison with the previous POWHEG generator}
\begin{figure}[!htb]
  \begin{center}
    \includegraphics[width=0.48\textwidth]{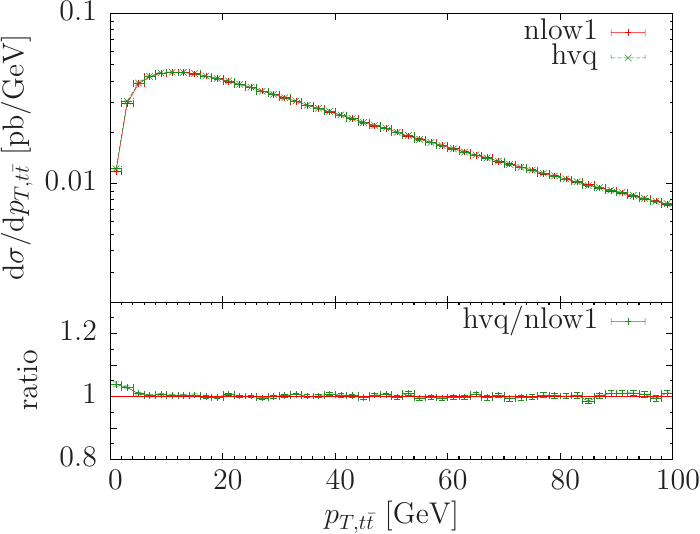}\hspace{0.4cm}
    \includegraphics[width=0.48\textwidth]{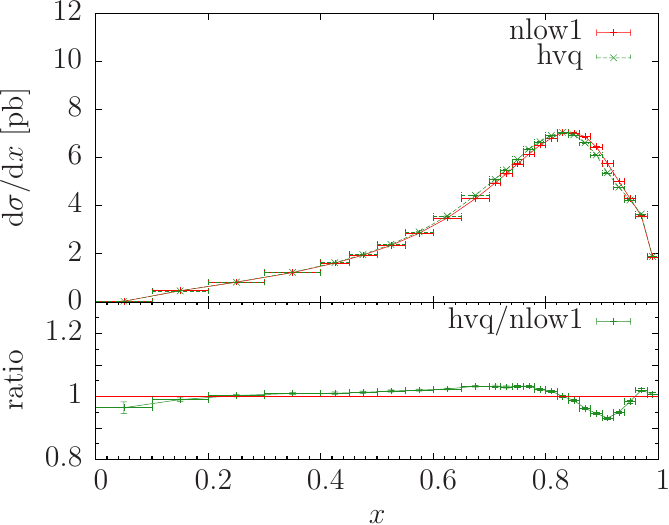}
  \end{center}
  \caption{Comparison of our ``nlow1'' generator and the {\tt hvq}
    generator for the $t\bar{t}$ pair transverse momentum (left) and
    for the $B$ meson fragmentation function (right). The {\tt hvq}
    result is multiplied by a factor of $1.05$, to match the
    normalization of the ``nlo'' one. No cuts have been imposed to
    obtain these plots.}
  \label{fig:ptpair-old-new}
\end{figure}

A \POWHEG{} generator for heavy quark production, capable to
describe $t\bar{t}$ production and decays, but including neither
exact LO spin correlations nor radiative corrections to decays,
was presented long ago~\cite{Frixione:2007nw}, and is now implemented
in the \POWHEGBOX{} framework. Here we compare our new generator with
the old one, that we will refer to as the \verb!hvq! generator.
We will limit the comparison to the ``nlow1'' mode
of the new generator, since it is the closest to the  \verb!hvq! one.
The differences between the ``nlow1'' and the other
modes have been studied in the previous sections,
so no other comparisons are needed.
The new ``nlow1'' generator differs from the \verb!hvq! one only in the
fact that the latter implements spin correlations in an approximate way,
according to the method presented in~\cite{Frixione:2007zp}.
We will essentially address two questions:
\begin{itemize}
\item Whether there are observable differences between the new
generator and the old generator for quantities that do not depend
upon the direction of the top or anti-top decay products (\emph{i.e.} that
are not sensitive to spin correlations).
\item Whether there are differences for observables that are sensitive
to spin correlations.
\end{itemize}
As  far as  the first  question is  concerned, we  have found, among
the large set of observables that we have examined,  small
differences in  the transverse momentum distribution  of the $t\bar{t}$
pair and in the bottom fragmentation function, as can be seen
in fig.~\ref{fig:ptpair-old-new}.
The total cross section in the ``nlow1'' and \verb!hvq! case differ,
the ``nlow1'' one being larger by about 5\%{}. This can be attributed
to the fact that other production mechanisms can act in the ``nlow1''
case, yielding a larger cross section below the $2m_{\rm top}$
threshold for the invariant mass of the $t\bar{t}$ pair.  The
transverse momentum difference is visible below 5\%{}, and it affects
the transverse momentum of the pair up to 4 GeV. It must be attributed
to the different way in which radiation is produced in the two
generators.  It is clearly a minor effect, not likely to affect
significantly realistic analysis.  The difference in the fragmentation
function appears as a light distortion in shape, for large values of
the $B$ energy fraction.  It is entirely due to \PYTHIAEight{}, since
in both the \verb!hvq! and ``nlow1'' generators radiation in decay is
handled by the shower generator.  We tentatively attribute this
difference to the fact that the slightly different radiation pattern
in production can influence strong radiation in the decay of coloured
resonances. We should in fact remember that the bottom quark is colour
connected (via the top quark) to particles arising from radiation in
production.  Notice that these differences in the fragmentation
function are much smaller than those found between ``allrad'' and
``nlow1'', due to the difference between an NLO exact and a shower
approximate treatment of radiation in decay.

\begin{figure}[!htb]
  \begin{center}
    \includegraphics[width=0.507\textwidth]{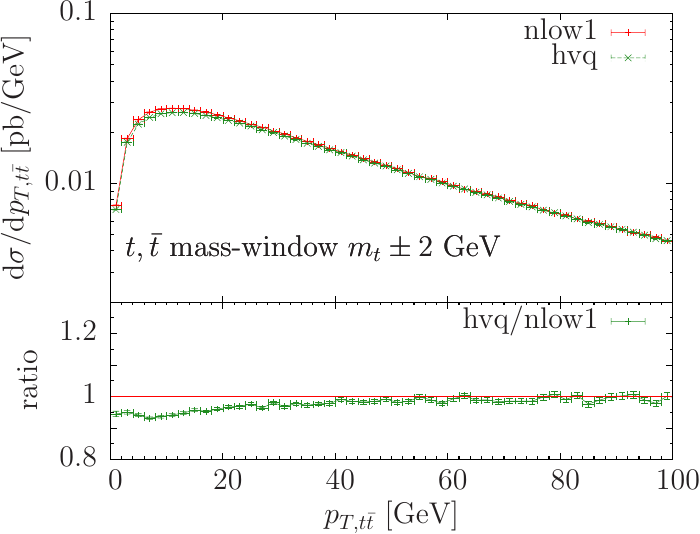}
  \end{center}
  \caption{Comparison of our ``nlow1'' (green) generator and the {\tt
      hvq} (red) generator for the $t\bar{t}$ pair transverse momentum
    when we require that the invariant mass of the top (and anti-top)
    decay products lie in a window of $\pm 2$ GeV around the top
    mass. No other cuts are imposed.}
  \label{fig:ptpair-old-new-masscut}
\end{figure}

In case severe cuts are imposed on the mass of the top decay products,
even larger differences are observed between the old and new generator.
In fig.~\ref{fig:ptpair-old-new-masscut}
we plot the transverse momentum of the $t\bar{t}$ pair when such cuts are imposed.
The relatively important difference is easily traced back to the way in which off-shell
effects are simulated in the {\tt hvq} generator. The event generation starts there with
an on-shell $t\bar{t}$ pair, produced according to the correct on-shell probability.
The momenta of the final state particles are then modified with a probability
proportional to the tree-level, off-shell matrix elements. This is done, however, without
changing the overall initial probability. The mapping of the on-shell to the off-shell
kinematics is obtained by first assigning new virtualities to the top and anti-top,
and then performing a reshuffling procedure, that does not
change the pair 3-momentum in the partonic CM frame.
Lower top and anti-top virtualities do however imply lower total CM energy, and are
therefore favoured, since they imply higher luminosity. A tight cut around the top mass window
will thus cut away these events, especially for low invariant mass of the  $t\bar{t}$ pair,
and for low value of the transverse momentum, where a reduced top virtuality has more impact
on the luminosities. Although it is unlikely that such features may actually spoil
realistic analysis, it is also true that they expose limitations of the {\tt hvq} generator
that are not present in the new implementation.

In order to study spin correlation effects, we have examined
the azimuthal distance of the two leptons, and the observables $\cos\theta_{l^+}$,
 $\cos\theta_{l^-}$. These are defined as the cosine of the angle between
the lepton in the top rest frame, with respect to the flight direction of the top
in the centre-of-mass frame of the $t\bar{t}$ system (see ref.~\cite{Bernreuther:2013aga}).
We have plotted $\cos\theta_{l^+}$ in ten bins of $\cos\theta_{l^-}$, and we have computed
the average of $\cos\theta_{l^+}\cos\theta_{l^-}$. We have seen no significant
differences among the ``nlow1'' and \verb!hvq! generators for any of these observables,
also when imposing the generic cuts that we have used in the present work.
A more detail comparison of the two generators for
observables sensitive to spin correlations is left to future work.

\section{Conclusions}\label{sec:Conc}
In this work we have presented an NLO accurate generator for $t\bar{t}$ production,
implemented in the \POWHEGBOX{} framework, where
radiative corrections in decays are also included. This generator
is based upon an NLO calculation performed in the narrow width approximation.
Finite width effects are however included in an approximate way. This generator improves
over a previous one~\cite{Frixione:2007nw} by the inclusion of exact spin correlations in decays, and by
the inclusion of NLO corrections to the decay processes.

We have compared the NLO-level output of our generator with the
calculation of ref.~\cite{Denner:2012yc}, where all graphs leading to the
final state for fully leptonic top decays
are included with all their interferences.  With
respect to this calculation, we reproduce exactly the leading order
result, and reproduce fairly the NLO one, provided one does not
look too far away from the resonance regions.

We have introduced three different methods for the approximate
implementation of off-shell effects. The least accurate one uses a
phase space mapping of off-shell configurations into on-shell ones,
and mimics the corresponding difference in the matrix element by a
simple Breit-Wigner reweighting (here dubbed BW method). The next
method (the DR method), reweights the matrix element using the ratio
of the exact LO double resonant matrix element divided by the on-shell
one, both computed with the kinematics of the underlying Born. Finally
our most accurate method is similar to the DR one, but makes use of
the full LO cross section for the given final state, including also
non-resonant contributions and interference effects.  We have found
that the DR and BW result are in fairly good agreement, thus giving us
confidence that our off-shell-on-shell mapping procedure is sound. The
full result is also in good agreement with the DR one, provided
that cuts suppressing the non-resonant $b\bar{b}$  production
via gluon splitting are applied.

We have studied in detail the effect of including NLO corrections
in the top decay, compared to letting the shower Monte Carlo handle
the decay process. By doing this, we have learned that it is better
to use an approach where production and decay are factorized, and we have
implemented such an approach in our generator. More specifically, since \POWHEG{}
normally generates only the hardest radiation, in a standard \POWHEG{}
implementation, only the hardest part of radiation in decay
would be corrected at the NLO level, while with our factorized prescription,
the hardest radiation in decays is always generated by \POWHEG{}.

We found that NLO corrections
in top decays mostly affect the $B$ meson fragmentation function,
and to a smaller extent observables that measure the hardness of
radiation in the top rest frame.
We are aware, however, of the fact the $B$ fragmentation properties
also depend upon Monte Carlo tuning. Ideally, one should fit the
parameters that affect $B$ fragmentation
with a \POWHEG{}+Shower setup for
 $e^+e^- \to b\bar{b}$, and then use the same fragmentation parameters
in the $t\bar{t}$ case.

We believe that our generator
will be particularly useful to discuss the systematics of top mass
determination from end point observables, like the lepton-$B$-meson
and lepton-$B$-jet invariant mass, and on similar observables that
make also use of the measured missing transverse energy.

We have also compared our new generator with the previous
\POWHEG{} one~\cite{Frixione:2007nw}, in order to see whether the better
treatment of off-shell effects and the inclusion of exact spin
correlations, leads to other important differences. As far as
spin correlations are concerned, we have not seen relevant
differences.
On the other hand, we have noticed that the previous implementation
of off-shell effects may lead to minor differences in the
transverse momentum distribution of the $t\bar{t}$ pair
in the small transverse momentum region, where we have reasons
to believe that the new implementation is more solid.

\appendix
\section{Interface to \PYTHIAEight{}}\label{app:pythiaIF}
Several issues have arisen with regard to the interface between \POWHEG{}
and \PYTHIAEight{}. In the following, we first discuss our relevant choices for the
settings, and then the entire procedure that we have used to implement our
radiation vetos in the various cases.
\subsection{\PYTHIAEight{} and \scalup{} settings}
Generally, \PYTHIAEight{} should comply with the requirement that 
radiation with hardness above the \scalup{} variable~\cite{Boos:2001cv}
should be vetoed. There is, however, an exception that does apply to our case.
According to the \PYTHIAEight{}
manual,\footnote{See \url{http://home.thep.lu.se/~torbjorn/pythia82html/Welcome.html}.}
if no light partons
(not arising from resonance decays) are present in the final state,
the \scalup{} variable is ignored. The \PYTHIAEight{} rational for this
behaviour is that, since no radiation is present at the LH level, there is
no danger of overcounting by allowing radiation in the full kinematically
allowed region. In our case, and especially in the ``nlow0'' case,
radiation in resonance decays and production do compete, and it is therefore
necessary to veto ISR even if no other radiation in production is present at the LH level.
This can be achieved in \PYTHIAEight{} with the calls
\begin{verbatim}
    pythia.readString("SpaceShower:pTmaxMatch = 1")
    pythia.readString("TimeShower:pTmaxMatch = 1"),
\end{verbatim}
that force \PYTHIAEight{} to blindly honour the \scalup{} requests.
All our \PYTHIAEight{} runs have been performed with the above settings.
\subsection{Implementation of the radiation vetos in resonance decays.}
Although \PYTHIAEight{} does implement a method for vetoing radiation in resonance
decays, we have preferred a different approach that allows us to have better
control. Our hardness definition in case of radiation
from $b$ quarks in $t$ decays, is given by
\begin{equation}\label{eq:hardnessbg}
  2 p_b\cdot p_g \frac{E_g}{E_b}=E_g^2\; 2(1-\cos \theta_{bg}),
\end{equation}
in the top rest frame,
that corresponds to the transverse momentum of the gluon
$E_g^2 \theta_{bg}^2$ in the soft-collinear limit. We found evidence that this
definition does not match exactly the one used by \PYTHIAEight{}.
We thus do the following. We thus look at the last top in the \PYTHIAEight{}
event record. This corresponds to the top after any (production stage) radiation,
and before its decay. \PYTHIAEight{} provides pointers to the first and
last particle in the shower record coming from the top decay. The list of
decay products can consist of the following:
\begin{itemize}
\item
Two particles: a $W$ and a $b$.
This happens if only a $W$ and a $b$ where present as decay products of the top
in the Les Houches interface.
In this case we follow the $b$ sons in the shower history. We have two cases:
\begin{itemize}
\item
The $b$ is pointing to a list containing a $b$ and a gluon. In this case
the scale of the shower radiation in the top decay, that we call \verb!St!,
 is computed applying formula
\ref{eq:hardnessbg} to this last $bg$ pair.
\item
The $b$ is pointing to a list of a single $b$ quark. In this case \verb!St! is set to a negative number.
\end{itemize}
\item
Three particles: $W$, $b$ and $g$.
This happens if a $W$, a $b$ and a gluon were present as decay products of the top
in the Les Houches interface (\emph{i.e.} \POWHEG{} has generated the hardest radiation).
The following cases may happen:
\begin{itemize}
\item 
The $b$ is pointing to a list containing a $b$ and a gluon. In this case
the scale of the shower radiation from the b in the top decay, \verb!Stb! is computed applying formula
\ref{eq:hardnessbg} to this last $bg$ pair.
\item
The $b$ is pointing to a list of a single $b$ quark. In this case \verb!Stb!
is set to a negative number.
\end{itemize}
As far as the gluon is concerned, we also have two cases:
\begin{itemize}
\item 
The gluon is pointing to a list containing two light partons 1 and 2. In this case
the scale of the shower radiation from the gluon in the top decay \verb!Stg! is computed applying formula:
\begin{equation}
  2 p_1\cdot p_2 \frac{E_1E_2}{E_1^2+E_2^2},
\end{equation}
again corresponding to the square of the transverse momentum of the softest particle
in the soft-collinear limit.
\item
The gluon is pointing to a list containing a single gluon.  In this
case \verb!Stg! is set to a negative value.
\end{itemize}
We now set \verb!St=max(Stb,Stg)!.
\end{itemize}
We apply this procedure to both the top and the anti-top, obtaining two scales, that
we call \verb!St+! and \verb!St-!.
We use them as follows, in our different running modes:
\begin{itemize}
\item
``allrad'': In this case we also compute the hardness of radiation in top and anti-top decay at the
Les Houches level, \verb!Pst+! and \verb!Pst-! respectively. We veto the event if either
\verb!St+ > Pst+! or \verb!St- > Pst-!.
\item
``nlow0'': In this case, we veto the event if either \verb!St+ > scalup! or \verb!St- > scalup!.
\item
``nlow1'': we never veto. In this case, the shower builds the resonance radiation according to its
own algorithms, with no restrictions. Radiation in production is instead vetoed internally
by the shower itself, according to the value of \scalup{}.
\item
``nlow4'': In this case, we veto the event if \verb!St+ > scalup!. The radiation from the anti-top is
handled by the shower. We take care to store the initial \scalup{} value, to be used in vetoing
radiation in $t$ decays, and to set it to the mass of the $t\bar{t}$ pair, in such a way that
radiation in production is handled by the shower alone.
\end{itemize}
When we veto an event, we discard it and re-run the shower with the same
Les Houches event. In other words,
we keep showering the same Les Houches event until all our conditions are met.
\section*{Acknowledgements}
We thank S. Kallweit and S. Pozzorini for sending us their numerical
 results. We also thank T. Sj\"ostrand and P. Skands for useful
 discussion about the $B$-meson fragmentation function in
 \PYTHIAEight{}. ER acknowledges support from the Munich Institute for Astro-
and Particle Physics (MIAPP) of the DFG cluster of excellence "Origin
and Structure of the Universe.  PN and ER also thank the Galileo
Galilei Institute for Theoretical Physics for hospitality and the INFN
for partial support during the completion of this work.
The research of JMC and RKE is supported by the US DOE under contract DE-AC02-07CH11359.
\bibliographystyle{JHEP}
\bibliography{paper}

\providecommand{\href}[2]{#2}\begingroup\raggedright\begin{thebibliography}{10}

\bibitem{Chatrchyan:2013ual}
{\bf CMS Collaboration} Collaboration, S.~Chatrchyan et~al., {\it {Measurement
  of the $t\bar{t}$ production cross section in the all-jet final state in pp
  collisions at $\sqrt{s}$ = 7 TeV}},  {\em JHEP} {\bf 1305} (2013) 065,
  [\href{http://xxx.lanl.gov/abs/1302.0508}{{\tt arXiv:1302.0508}}].

\bibitem{Chatrchyan:2012ria}
{\bf CMS Collaboration} Collaboration, S.~Chatrchyan et~al., {\it {Measurement
  of the $t\bar{t}$ production cross section in $pp$ collisions at $\sqrt{s}=7$
  TeV with lepton + jets final states}},  {\em Phys.Lett.} {\bf B720} (2013)
  83--104, [\href{http://xxx.lanl.gov/abs/1212.6682}{{\tt arXiv:1212.6682}}].

\bibitem{Chatrchyan:2013faa}
{\bf CMS Collaboration} Collaboration, S.~Chatrchyan et~al., {\it {Measurement
  of the $t \bar{t}$ production cross section in the dilepton channel in pp
  collisions at $\sqrt{s}$ = 8 TeV}},  {\em JHEP} {\bf 1402} (2014) 024,
  [\href{http://xxx.lanl.gov/abs/1312.7582}{{\tt arXiv:1312.7582}}].

\bibitem{Aad:2014iaa}
{\bf ATLAS Collaboration} Collaboration, G.~Aad et~al., {\it {Measurement of
  the $t\bar{t}$ production cross-section as a function of jet multiplicity and
  jet transverse momentum in 7 TeV proton-proton collisions with the ATLAS
  detector}},  \href{http://xxx.lanl.gov/abs/1407.0891}{{\tt arXiv:1407.0891}}.

\bibitem{Aad:2014kva}
{\bf ATLAS Collaboration} Collaboration, G.~Aad et~al., {\it {Measurement of
  the $t\bar{t}$ production cross-section using $e\mu$ events with $b$-tagged
  jets in $pp$ collisions at $\sqrt{s}=7$ and 8 TeV with the ATLAS detector}},
  \href{http://xxx.lanl.gov/abs/1406.5375}{{\tt arXiv:1406.5375}}.

\bibitem{Aad:2012vip}
{\bf ATLAS Collaboration} Collaboration, G.~Aad et~al., {\it {Measurement of
  the ttbar production cross section in the tau+jets channel using the ATLAS
  detector}},  {\em Eur.Phys.J.} {\bf C73} (2013) 2328,
  [\href{http://xxx.lanl.gov/abs/1211.7205}{{\tt arXiv:1211.7205}}].

\bibitem{Chatrchyan:2013xza}
{\bf CMS Collaboration} Collaboration, S.~Chatrchyan et~al., {\it {Measurement
  of the top-quark mass in all-jets $t\bar{t}$ events in pp collisions at
  $\sqrt{s}$=7 TeV}},  {\em Eur.Phys.J.} {\bf C74} (2014) 2758,
  [\href{http://xxx.lanl.gov/abs/1307.4617}{{\tt arXiv:1307.4617}}].

\bibitem{Chatrchyan:2013haa}
{\bf CMS Collaboration} Collaboration, S.~Chatrchyan et~al., {\it
  {Determination of the top-quark pole mass and strong coupling constant from
  the ttbar production cross section in pp collisions at sqrt(s) = 7 TeV}},
  {\em Phys.Lett.} {\bf B728} (2014) 496--517,
  [\href{http://xxx.lanl.gov/abs/1307.1907}{{\tt arXiv:1307.1907}}].

\bibitem{Chatrchyan:2013boa}
{\bf CMS Collaboration} Collaboration, S.~Chatrchyan et~al., {\it {Measurement
  of masses in the $t \bar{t}$ system by kinematic endpoints in pp collisions
  at $\sqrt{s}$ = 7 TeV}},  {\em Eur.Phys.J.} {\bf C73} (2013) 2494,
  [\href{http://xxx.lanl.gov/abs/1304.5783}{{\tt arXiv:1304.5783}}].

\bibitem{Chatrchyan:2013wua}
{\bf CMS Collaboration} Collaboration, S.~Chatrchyan et~al., {\it {Measurements
  of $t\bar{t}$ spin correlations and top-quark polarization using dilepton
  final states in pp collisions at $\sqrt{s}$ = 7 TeV}},  {\em Phys.Rev.Lett.}
  {\bf 112} (2014) 182001, [\href{http://xxx.lanl.gov/abs/1311.3924}{{\tt
  arXiv:1311.3924}}].

\bibitem{Chatrchyan:2012saa}
{\bf CMS Collaboration} Collaboration, S.~Chatrchyan et~al., {\it {Measurement
  of differential top-quark pair production cross sections in $pp$ colisions at
  $\sqrt{s}=7$ TeV}},  {\em Eur.Phys.J.} {\bf C73} (2013) 2339,
  [\href{http://xxx.lanl.gov/abs/1211.2220}{{\tt arXiv:1211.2220}}].

\bibitem{Aad:2014zka}
{\bf ATLAS Collaboration} Collaboration, G.~Aad et~al., {\it {Measurements of
  normalized differential cross-sections for ttbar production in pp collisions
  at sqrt(s) = 7 TeV using the ATLAS detector}},
  \href{http://xxx.lanl.gov/abs/1407.0371}{{\tt arXiv:1407.0371}}.

\bibitem{Aad:2012hg}
{\bf ATLAS Collaboration} Collaboration, G.~Aad et~al., {\it {Measurements of
  top quark pair relative differential cross-sections with ATLAS in $pp$
  collisions at $\sqrt{s}=7$ TeV}},  {\em Eur.Phys.J.} {\bf C73} (2013) 2261,
  [\href{http://xxx.lanl.gov/abs/1207.5644}{{\tt arXiv:1207.5644}}].

\bibitem{Czakon:2013goa}
M.~Czakon, P.~Fiedler, and A.~Mitov, {\it {Total Top-Quark Pair-Production
  Cross Section at Hadron Colliders Through $O(α\frac{4}{S})$}},  {\em
  Phys.Rev.Lett.} {\bf 110} (2013), no.~25 252004,
  [\href{http://xxx.lanl.gov/abs/1303.6254}{{\tt arXiv:1303.6254}}].

\bibitem{Abelof:2014fza}
G.~Abelof, A.~Gehrmann-De~Ridder, P.~Maierhofer, and S.~Pozzorini, {\it {NNLO
  QCD subtraction for top-antitop production in the $ q\overline{q} $
  channel}},  {\em JHEP} {\bf 1408} (2014) 035,
  [\href{http://xxx.lanl.gov/abs/1404.6493}{{\tt arXiv:1404.6493}}].

\bibitem{Czakon:2014xsa}
M.~Czakon, P.~Fiedler, and A.~Mitov, {\it {Resolving the Tevatron top quark
  forward-backward asymmetry puzzle}},
  \href{http://xxx.lanl.gov/abs/1411.3007}{{\tt arXiv:1411.3007}}.

\bibitem{Brucherseifer:2013iv}
M.~Brucherseifer, F.~Caola, and K.~Melnikov, {\it {$\mathcal O(\alpha_s^2)$
  corrections to fully-differential top quark decays}},  {\em JHEP} {\bf 1304}
  (2013) 059, [\href{http://xxx.lanl.gov/abs/1301.7133}{{\tt
  arXiv:1301.7133}}].

\bibitem{Mangano:1991jk}
M.~L. Mangano, P.~Nason, and G.~Ridolfi, {\it {Heavy quark correlations in
  hadron collisions at next-to-leading order}},  {\em Nucl.Phys.} {\bf B373}
  (1992) 295--345.

\bibitem{Frixione:2003ei}
S.~Frixione, P.~Nason, and B.~R. Webber, {\it {Matching NLO QCD and parton
  showers in heavy flavor production}},  {\em JHEP} {\bf 0308} (2003) 007,
  [\href{http://xxx.lanl.gov/abs/hep-ph/0305252}{{\tt hep-ph/0305252}}].

\bibitem{Frixione:2007nw}
S.~Frixione, P.~Nason, and G.~Ridolfi, {\it {A Positive-weight
  next-to-leading-order Monte Carlo for heavy flavour hadroproduction}},  {\em
  JHEP} {\bf 0709} (2007) 126, [\href{http://xxx.lanl.gov/abs/0707.3088}{{\tt
  arXiv:0707.3088}}].

\bibitem{Frixione:2007zp}
S.~Frixione, E.~Laenen, P.~Motylinski, and B.~R. Webber, {\it {Angular
  correlations of lepton pairs from vector boson and top quark decays in Monte
  Carlo simulations}},  {\em JHEP} {\bf 0704} (2007) 081,
  [\href{http://xxx.lanl.gov/abs/hep-ph/0702198}{{\tt hep-ph/0702198}}].

\bibitem{Bernreuther:2004jv}
W.~Bernreuther, A.~Brandenburg, Z.~Si, and P.~Uwer, {\it {Top quark pair
  production and decay at hadron colliders}},  {\em Nucl.Phys.} {\bf B690}
  (2004) 81--137, [\href{http://xxx.lanl.gov/abs/hep-ph/0403035}{{\tt
  hep-ph/0403035}}].

\bibitem{Melnikov:2009dn}
K.~Melnikov and M.~Schulze, {\it {NLO QCD corrections to top quark pair
  production and decay at hadron colliders}},  {\em JHEP} {\bf 0908} (2009)
  049, [\href{http://xxx.lanl.gov/abs/0907.3090}{{\tt arXiv:0907.3090}}].

\bibitem{Campbell:2012uf}
J.~M. Campbell and R.~K. Ellis, {\it {Top-quark processes at NLO in production
  and decay}},  {\em J.Phys.G: Nucl. Part. Phys.} {\bf 42} (2015) 015005,
  [\href{http://xxx.lanl.gov/abs/1204.1513}{{\tt arXiv:1204.1513}}].

\bibitem{Denner:2010jp}
A.~Denner, S.~Dittmaier, S.~Kallweit, and S.~Pozzorini, {\it {NLO QCD
  corrections to WWbb production at hadron colliders}},  {\em Phys.Rev.Lett.}
  {\bf 106} (2011) 052001, [\href{http://xxx.lanl.gov/abs/1012.3975}{{\tt
  arXiv:1012.3975}}].

\bibitem{Denner:2012yc}
A.~Denner, S.~Dittmaier, S.~Kallweit, and S.~Pozzorini, {\it {NLO QCD
  corrections to off-shell top-antitop production with leptonic decays at
  hadron colliders}},  {\em JHEP} {\bf 1210} (2012) 110,
  [\href{http://xxx.lanl.gov/abs/1207.5018}{{\tt arXiv:1207.5018}}].

\bibitem{Bevilacqua:2010qb}
G.~Bevilacqua, M.~Czakon, A.~van Hameren, C.~G. Papadopoulos, and M.~Worek,
  {\it {Complete off-shell effects in top quark pair hadroproduction with
  leptonic decay at next-to-leading order}},  {\em JHEP} {\bf 1102} (2011) 083,
  [\href{http://xxx.lanl.gov/abs/1012.4230}{{\tt arXiv:1012.4230}}].

\bibitem{Heinrich:2013qaa}
G.~Heinrich, A.~Maier, R.~Nisius, J.~Schlenk, and J.~Winter, {\it {NLO QCD
  corrections to $W^{+} W^{-}b\bar{b}$ production with leptonic decays in the
  light of top quark mass and asymmetry measurements}},  {\em JHEP} {\bf 1406}
  (2014) 158, [\href{http://xxx.lanl.gov/abs/1312.6659}{{\tt
  arXiv:1312.6659}}].

\bibitem{Frederix:2013gra}
R.~Frederix, {\it {The top induced backgrounds to Higgs production in the WW
  $\to$ llvv decay channel at NLO in QCD}},  {\em Phys.Rev.Lett.} {\bf 112}
  (2014) 082002, [\href{http://xxx.lanl.gov/abs/1311.4893}{{\tt
  arXiv:1311.4893}}].

\bibitem{Cascioli:2013wga}
F.~Cascioli, S.~Kallweit, P.~Maierhöfer, and S.~Pozzorini, {\it {A unified NLO
  description of top-pair and associated Wt production}},  {\em Eur.Phys.J.}
  {\bf C74} (2014) 2783, [\href{http://xxx.lanl.gov/abs/1312.0546}{{\tt
  arXiv:1312.0546}}].

\bibitem{Garzelli:2014dka}
M.~V. Garzelli, A.~Kardos, and Z.~Trócsanyi, {\it {Hadroproduction of $W^{+}
  W^{-} b \bar{b}$ at NLO accuracy matched with shower Monte Carlo programs}},
  {\em JHEP} {\bf 1408} (2014) 069,
  [\href{http://xxx.lanl.gov/abs/1405.5859}{{\tt arXiv:1405.5859}}].

\bibitem{Nason:2004rx}
P.~Nason, {\it {A New method for combining NLO QCD with shower Monte Carlo
  algorithms}},  {\em JHEP} {\bf 0411} (2004) 040,
  [\href{http://xxx.lanl.gov/abs/hep-ph/0409146}{{\tt hep-ph/0409146}}].

\bibitem{Frixione:2002ik}
S.~Frixione and B.~R. Webber, {\it {Matching NLO QCD computations and parton
  shower simulations}},  {\em JHEP} {\bf 0206} (2002) 029,
  [\href{http://xxx.lanl.gov/abs/hep-ph/0204244}{{\tt hep-ph/0204244}}].

\bibitem{Badger:2011yu}
S.~Badger, R.~Sattler, and V.~Yundin, {\it {One-Loop Helicity Amplitudes for
  $t\bar{t}$ Production at Hadron Colliders}},  {\em Phys.Rev.} {\bf D83}
  (2011) 074020, [\href{http://xxx.lanl.gov/abs/1101.5947}{{\tt
  arXiv:1101.5947}}].

\bibitem{Alioli:2010xd}
S.~Alioli, P.~Nason, C.~Oleari, and E.~Re, {\it {A general framework for
  implementing NLO calculations in shower Monte Carlo programs: the POWHEG
  BOX}},  {\em JHEP} {\bf 1006} (2010) 043,
  [\href{http://xxx.lanl.gov/abs/1002.2581}{{\tt arXiv:1002.2581}}].

\bibitem{Alwall:2011uj}
J.~Alwall, M.~Herquet, F.~Maltoni, O.~Mattelaer, and T.~Stelzer, {\it {MadGraph
  5 : Going Beyond}},  {\em JHEP} {\bf 1106} (2011) 128,
  [\href{http://xxx.lanl.gov/abs/1106.0522}{{\tt arXiv:1106.0522}}].

\bibitem{Boos:2001cv}
E.~Boos, M.~Dobbs, W.~Giele, I.~Hinchliffe, J.~Huston, et~al., {\it {Generic
  user process interface for event generators}},
  \href{http://xxx.lanl.gov/abs/hep-ph/0109068}{{\tt hep-ph/0109068}}.

\bibitem{Alioli:2010qp}
S.~Alioli, P.~Nason, C.~Oleari, and E.~Re, {\it {Vector boson plus one jet
  production in POWHEG}},  {\em JHEP} {\bf 1101} (2011) 095,
  [\href{http://xxx.lanl.gov/abs/1009.5594}{{\tt arXiv:1009.5594}}].

\bibitem{Kardos:2014dua}
A.~Kardos, P.~Nason, and C.~Oleari, {\it {Three-jet production in POWHEG}},
  {\em JHEP} {\bf 1404} (2014) 043,
  [\href{http://xxx.lanl.gov/abs/1402.4001}{{\tt arXiv:1402.4001}}].

\bibitem{Martin:2009iq}
A.~Martin, W.~Stirling, R.~Thorne, and G.~Watt, {\it {Parton distributions for
  the LHC}},  {\em Eur.Phys.J.} {\bf C63} (2009) 189--285,
  [\href{http://xxx.lanl.gov/abs/0901.0002}{{\tt arXiv:0901.0002}}].

\bibitem{Lai:2010vv}
H.-L. Lai, M.~Guzzi, J.~Huston, Z.~Li, P.~M. Nadolsky, et~al., {\it {New parton
  distributions for collider physics}},  {\em Phys.Rev.} {\bf D82} (2010)
  074024, [\href{http://xxx.lanl.gov/abs/1007.2241}{{\tt arXiv:1007.2241}}].

\bibitem{Ball:2010de}
R.~D. Ball, L.~Del~Debbio, S.~Forte, A.~Guffanti, J.~I. Latorre, et~al., {\it
  {A first unbiased global NLO determination of parton distributions and their
  uncertainties}},  {\em Nucl.Phys.} {\bf B838} (2010) 136--206,
  [\href{http://xxx.lanl.gov/abs/1002.4407}{{\tt arXiv:1002.4407}}].

\bibitem{Cacciari:2008gp}
M.~Cacciari, G.~P. Salam, and G.~Soyez, {\it {The Anti-k(t) jet clustering
  algorithm}},  {\em JHEP} {\bf 0804} (2008) 063,
  [\href{http://xxx.lanl.gov/abs/0802.1189}{{\tt arXiv:0802.1189}}].

\bibitem{Cacciari:2011ma}
M.~Cacciari, G.~P. Salam, and G.~Soyez, {\it {FastJet User Manual}},  {\em
  Eur.Phys.J.} {\bf C72} (2012) 1896,
  [\href{http://xxx.lanl.gov/abs/1111.6097}{{\tt arXiv:1111.6097}}].

\bibitem{Bernreuther:2013aga}
W.~Bernreuther and Z.-G. Si, {\it {Top quark spin correlations and polarization
  at the LHC: standard model predictions and effects of anomalous top chromo
  moments}},  {\em Phys.Lett.} {\bf B725} (2013), no.~1-3 115--122,
  [\href{http://xxx.lanl.gov/abs/1305.2066}{{\tt arXiv:1305.2066}}].

\end{thebibliography}\endgroup

\end{document}